\documentclass[preprint,12pt]{elsarticle}

\usepackage{amssymb}
\usepackage{float}
\usepackage{amsmath}
\usepackage{amsthm}
\usepackage{todonotes}
\usepackage{xurl}
\usepackage[ruled,linesnumbered, vlined]{algorithm2e}
\usepackage{xcolor}
\journal{Data \& Knowledge Engineering}

\newcommand{\hips}[0]{\textsc{HFips}}
\newcommand{\fips}[0]{\textsc{Fips}}

\newcommand{\alea}[0]{\textsc{Uniform}}
\newcommand{\NIP}[0]{NIP}

\newcommand{\IH}[1]{IPH(#1)}
\newcommand{\TIL}[2]{TIL_{#1}(#2)}
\newcommand{\I}[1]{I(#1)}
\newcommand{\J}[1]{J(#1)}

\newtheorem{property}{Property}

\newtheorem{example}{Example}

\let\oldnl\nl
\newcommand{\nonl}{\renewcommand{\nl}{\let\nl\oldnl}}

\def\N{\mathcal{N}} 
\def\M{\mathcal{M}} 
\def\im{{m}} 
\def\G{\mathcal{G}} 
\def\ig{{g}} 
\def\W{\mathcal{N}} 
\def\IP{\mathcal{V}} 
\def\gAsIP{\mathcal{B}}
\def\cover{{\it cover}} 
\def\freq{{\it freq}} 
\def\vol{{\it vol}} 

\newcommand{\inff}[1]{\underline{#1}}
\newcommand{\supp}[1]{\overline{#1}}

\def\lang{\mathcal{L}} 

\newcommand{\val}[2]{v_{\ig^{#1},\im^{#2}}}

\def\cover{{\it cover}} 
\def\freq{{\it freq}} 
\def\desc{{\it desc}}

\begin{document}

\begin{frontmatter}



\title{
Efficiently Sampling Interval Patterns\\ from Numerical Databases}


\author[unicaen]{Djawad Bekkoucha} 
\author[EPITA]{Lamine Diop}
\author[unicaen]{Abdelkader Ouali}
\author[unicaen]{Bruno Crémilleux}
\author[unicaen]{Patrice Boizumault}

\affiliation[unicaen]{
    organization={Université de Caen Normandie, ENSICAEN, CNRS, Normandie Univ, GREYC UMR6072}, 
    addressline={6 Boulevard Marechal Juin}, 
    city={Caen}, 
    postcode={F-14000}, 
    country={France}
}

\affiliation[EPITA]{
    organization={EPITA Research Laboratory (LRE)}, 
    addressline={Le Kremlin-Bicetre}, 
    city={Paris}, 
    postcode={FR-94276}, 
    country={France}
}

\begin{abstract}

Pattern sampling has emerged as a promising approach for information discovery in large databases, allowing analysts to focus on a manageable subset of patterns. In this approach, patterns are randomly drawn based on an interestingness measure, such as frequency or hyper-volume. This paper presents the first sampling approach designed to handle interval patterns in numerical databases. This approach, named \fips{}, samples interval patterns proportionally to their frequency. It uses a multi-step sampling procedure and addresses a key challenge in numerical data: accurately determining the number of interval patterns that cover each object. We extend this work with \hips{}, which samples interval patterns proportionally to both their frequency and hyper-volume. These methods efficiently tackle the well-known long-tail phenomenon in pattern sampling. We formally prove that \fips{} and \hips{} sample interval patterns in proportion to their frequency and the product of hyper-volume and frequency, respectively. Through experiments on several databases, we demonstrate the quality of the obtained patterns and their robustness against the long-tail phenomenon.

\end{abstract}




\end{frontmatter}

\section{Introduction}

Data scientists have a central role for knowledge discovery from data.  
In practice, analysts want to interact (visualise, select, explore) not only with the data, but also with the patterns or models supported by the data. To carry out such processes, it is essential to produce high quality results within a very short time so that the analyst does not have to wait. Otherwise, there is a risk that they will disconnect from the system. Pattern sampling is one solution to this challenge~\cite{DBLP:conf/pakdd/DzyubaL17}.

Pattern sampling aims at randomly selecting a pattern with a probability proportional to an interestingness measure~\cite{DBLP:journals/pvldb/alhassan2009, DBLP:journals/datamine/DzyubaLR17, DBLP:conf/kdd/BoleyLPG11} , such as frequency. For example, a pattern $\IP_1$ that is twice as frequent as a pattern $\IP_2$ will be twice as likely to be selected.
A naive approach to pattern sampling is to generate the entire set of patterns and then perform a weighted draw based on the interestingness measure. However, in practice, this fails due to the large size of the search space, even for simple pattern languages such as itemsets, i.e. data described by Boolean values. An answer to this problem was provided by Boley et al.~\cite{DBLP:conf/kdd/BoleyLPG11}, who defined an approach based on two successive draws. Once the problem has been properly decomposed, this approach guarantees an exact draw proportionally to the distribution resulting from the interestingness measure. Itemsets are the patterns considered in~\cite{DBLP:conf/kdd/BoleyLPG11}.

Numerical data are present in a wide range of applications such as transcriptome analysis~\cite{DBLP:journals/isb/BlachonPBRBG07}, medicine~\cite{DBLP:journals/isb/BlachonPBRBG07} and energy~\cite{DBLP:conf/dsaa/TschoraGPPR23}.
Even if handling numerical data to define patterns goes back to the origins of data mining~\cite{DBLP:conf/sigmod/SrikantA96}, pattern mining in numerical data remains a challenging task. A simple approach to cope with numerical data is to reuse existing methods for discrete data by ﬁrst converting data into a binary representation~\cite{Dougherty95supervisedand}.  However, it is well-known that the binarization process often leads to a loss of information.

Kaytoue at al.~\cite{kaytoue-ijcai11} introduced interval patterns, which preserve all the original information. However, this framework faces scalability issues due to the large number of generated patterns, which highlights the importance of pattern sampling in this context. To the best of our knowledge, there is only one sampling method designed for numerical data~\cite{DBLP:conf/sdm/GiacomettiS18}. This method uses a metric to construct neighbourhood patterns whose relevance is characterized by a density measure. Data are considered continuous and the method requires the setting of a parameter that defines the size (diameter) of a pattern.

{Our work} focuses on sampling in numerical data. We {strive} to keep the complete original information expressed by the numerical data by using \textit{Interval Patterns}~\cite{kaytoue-ijcai11}. We present \fips{} and \hips{}, two exact and non-enumerative methods for {sampling interval patterns}. \fips{} samples interval patterns proportionally to their frequency, while \hips{} extends this by incorporating hyper-volume, resulting in patterns sampled proportionally to the product of their frequency and hyper-volume. The key challenge for both methods is to compute, for each object in the database, its contribution to the overall sampling distribution.  For example, in the case of the frequency, the main challenge is to determine the exact number of interval patterns covering an object and without enumerating all covering patterns.
This allows us to bias the sampling toward patterns with high interestingness scores. We formally prove that \fips{} and \hips{} sample patterns proportionally to the desired distributions. Experimentally, we evaluate the quality of the obtained patterns by using several criteria (frequency, hyper-volume times frequency, impact of the long tail phenomenon, diversity, speed, and plausibility). \fips{} and \hips{} are the first methods for interval pattern sampling.

The paper is organized as follows. Section~\ref{sec:context} introduces the notations, definitions, and the problem statement. Section~\ref{sec:rw} discusses related work on pattern sampling. Section~\ref{sec:contrib} presents the \fips{} method, while Section~\ref{contrib:HFIPS} extends it with \hips{}. Finally, Section~\ref{sec:exp} reports an experimental evaluation of the proposed approaches.

\section{Preliminaries} 
\label{sec:context}



\subsection{{Numerical Dataset}}
A {\it numerical dataset} $\N$ is defined by
a set of objects $\G$ where each object is described by a set of attributes $\M$.
Each attribute $m \in \M$ has a range $\W_m$ which {is a finite set containing all the values of the data occurring in attribute $m$}. 
 An object $\ig \in  \G$ is defined by a vector of numerical values  $< \val{}{} >_{\forall \im \in\M}$.
A dataset where the values of all attributes are binary $\W_m=\{0,1\}, \forall m \in \M$, is a special case of a numerical dataset and referred as a {\it binary dataset}. 

\begin{example}
Table~\ref{dataset:nv:exemple} shows a running example of a numerical dataset containing 5~objects $\G=\{g_1,g_2, g_3, g_4, g_5\}$, 
each object is described by 3~attributes $\M=\{m_1,m_2, m_3\}$.     
\end{example}

\begin{table}[]
	\centering
	\scalebox{1}{
	\begin{tabular}{c c c c}
		& $m_1$ & $m_2$ & $m_3$ \\
		\hline
		$g_1$&2 &8 &130\\
		$g_2$&4 &12 &102\\
		$g_3$&3 &7 &91\\
		$g_4$&2 &9 &101\\
		$g_5$&6 &12 &110\\
	\end{tabular}
	}
	\caption{A running example of a numerical dataset $\N$}
	\label{dataset:nv:exemple}
\end{table}

\subsection{{Interval Patterns}}
Patterns in numerical datasets can be represented in many ways, we use the notion of Interval Pattern~\cite{kaytoue-ijcai11}
\label{def:intervalpatterns} 
which is defined as a vector of intervals $\IP =  \langle [a_m, b_m] \rangle_{\forall m \in \M}$, where $a_m, b_m \in \W_m$ and $a_m \leq b_m$.
Each dimension of the vector $\IP$ corresponds to an attribute following a canonical order on the set of attributes $\M$. We denote $\gAsIP[g] =  \langle [\val{}{}, \val{}{}] \rangle_{ \forall \im \in \M}$ as the vector of intervals corresponding to an object identified by $\ig$. An object $\ig$ is an occurrence of the interval pattern $\IP$ if each interval in the vector $\gAsIP[g]$ is included in the interval of $\IP$, i.e. $\gAsIP[g] \sqsubseteq \IP \iff [\val{}{}, \val{}{}] \subseteq [a_m, b_m], \forall \im \in \M$.
The cover of $\IP$ in $\N$ is the set of objects $\ig \in \G$ occurring in $\IP$, i.e. {$\cover(\IP)= \{\ig \in \mathcal{G} ~|~   \gAsIP[g] \sqsubseteq \IP\}$.}
 \label{def:cover}
 
\begin{example}
  {In the example dataset of Table \ref{dataset:nv:exemple}}, $\IP = \langle [3, 4], [7, 12], [91, 130] \rangle $ is an interval pattern covering the objects $\{g_2, g_3\}$. $\gAsIP[g_2]$$ =  $ $\langle [4, 4]$, $[12, 12],$ $[102,102] \rangle$ is the vector of intervals identified by the object $g_2$ and an occurrence of $\IP$.
\end{example}

\label{def:frequency}
The frequency of $\IP$ is the {cardinality} of its cover, i.e. $\freq(\IP)=|\cover(\IP)|$. 
Given a minimum frequency threshold $\theta$, the interval pattern $\IP$ is frequent if and only if $\freq(\IP) \geq \theta$.
\label{def:description} 
{The smallest description of a subset of objects $G \subseteq \G$ is the smallest interval pattern covering the set of objects $G$}. Formally the smallest description of $G$ is the interval pattern $\IP$ {such that for each} $\ig ~\in~ G$, $\ig$ is an occurrence of $\IP$, i.e. $\desc(G)= \langle[a_m,~b_m]\rangle_{ \forall \im \in \M} \text{ such that } a_m = min(\{\val{}{}~|~ g~ \in G \}) \text{ and } b_m = max(\{\val{}{}~|~ g~ \in G \})$. 
The hyper-volume of an interval pattern $\IP = \langle[a_m,~b_m]\rangle_{ \forall \im \in \M} $ corresponds to the product of the lengths of its intervals, i.e.
$\vol(\IP)= \prod_{[a_m,b_m] \in \IP} ( b_m - a_m )$. 

{Note that for an interval pattern $\IP$, if the interval associated to an attribute $m \in \M$ has its lower bound equal to the smallest possible value of $\W_m$ and its upper bound equal to the highest possible value of $\W_m$, then the interval is unconstrained. Otherwise, it is considered constrained on the subset of values it contains.}

Let $\mathcal{L}$ be the language of interval patterns, which corresponds to the set of all possible interval patterns. The size of the search space is the product of the total number of possible intervals for each attribute. Formally this is given by:
\[
\prod_{m \in \M}{\sum_{k=1}^{|\W_m|}{k}} = \prod_{m \in \M}{\frac{|\W_{m}| (|\W_{m}|+1)}{2}}
\]

Consider the dataset presented in Table~\ref{dataset:nv:exemple}, which contains 5 objects and 3 attributes. Each of the attributes $m_1$ and $m_2$ contains 4 distinct values, respectively $\W_{m_1}=\{2,3,4,6\}$ and $\W_{m_2}=\{7,8,9,12\}$, and attribute $m_3$ contains 5 distinct values, namely $\W_{m_3}=\{91,101,102,110,130\}$. The total number of interval patterns in the search space is given by :

\[
\underbrace{\frac{4 \times 5}{2} }_{m_1} \times \underbrace{\frac{4 \times 5}{2}}_{m_2} \times \underbrace{\frac{5 \times 6}{2}}_{m_3} = 1500 \text{ interval patterns}
\]

This example illustrates the size of the interval pattern search space. An exhaustive enumeration of all these patterns would be infeasible both in terms of time and memory. Moreover, given the large number of patterns, identifying the relevant ones would become a challenging task. This motivates the use of pattern sampling to explore the search space more efficiently.

\subsection{Problem Statement}

Let $\Omega$ be a population and $f : \Omega \longrightarrow [0,1]$ a measure. The notation $x \sim f(\Omega)$ {denotes} that the element $x$ is drawn randomly from $\Omega$ with a probability distribution $\pi(x) = f(x)/\sum_{x' \in \Omega} f(x')$.

Given a numerical database $\N$, the language of interval patterns $\mathcal{L}$ and $k \in \mathbb{N}$, the problem of sampling interval patterns {consists in} retrieving $k$ patterns $\IP_1,\ldots, \IP_k$  from $\mathcal{L}$, where each pattern $\IP_i$ is randomly drawn with replacement with a probability proportional to its value of $f$. 
Formally, this corresponds to:  
 \[ Sampling_{k}(\mathcal{L}, \N, f)= \bigcup_{i=1}^{k} ~\{\IP_i \in \mathcal{L}|~\IP_i \sim {f}(\IP_i)  \}\]

This work focuses on the frequency measure, {i.e. $\freq(\IP_{i})$}, and demonstrates how the principle can be extended to the product of hyper-volume and frequency, {i.e. $\vol(\IP_{i}) \times \freq(\IP_{i})$}.

\section{Related Work}
\label{sec:rw}

We have organized the overview of the pattern sampling methods into three families: probabilistic, declarative and multi-step methods.

The first family is based on Markov chain Monte Carlo algorithms, where the target sampling distribution {corresponds} to the stationary distribution of a random walk. These methods can handle various pattern languages and interestingness measures. For example, {Hasan and Zaki} \cite{DBLP:journals/pvldb/alhassan2009} introduced the first pattern sampling approach that {handles a graph pattern language} and is biased towards a frequency measure. Similarly, Boley et al.~\cite{DBLP:conf/sdm/BoleyGG10} proposed sampling formal concepts using strictly positive interestingness measures, while Bendimerad et al.~\cite{DBLP:conf/ida/BendimeradLPRB20} focused on sampling tiles proportionally to a subjective interest. Camelin et al. \cite{DBLP:journals/dke/CamelinLPT25} define a compression-based approach leveraging the LCM algorithm  \cite{DBLP:conf/fimi/UnoKA04} to sample frequent and closed itemsets. Finally, Opran et al.~\cite{RNTI/papers/1002998} introduced \textsc{SIMAS}, an interactive approach for sampling classification rules according to user-defined interest. Despite their accuracy, stochastic approaches often suffer from slow convergence.

The second family is based on declarative paradigms. \textsc{flexics}~\cite{DBLP:journals/datamine/DzyubaLR17} uses a SAT solver to sample itemsets and considers a wide range of measures. This approach randomly partitions the search space into different cells by recursively generating XOR constraints on variables associated with the items describing the patterns. However, modeling other pattern languages is challenging, as the encoding needs to be adapted for each language to ensure efficient implementation.

The third family is composed of multi-step methods. These methods partition the occurrences of the patterns into well-devised groups, enabling group sampling based on their weights, and then uniformly sampling a pattern within the selected group. The challenge is to find an appropriate multi-step decomposition to obtain the desired distribution. This approach was pioneered by Boley et al.~\cite{DBLP:conf/kdd/BoleyLPG11}, using a two-step draw to sample patterns according to the frequency. Diop et al.~\cite{DBLP:journals/kais/DiopDGLS20} extend this approach with a three-step method for sequential pattern sampling, using a frequency measure and a maximum length constraint. Soulet~\cite{DBLP:conf/f-egc/Soulet23} combines the sampling and constrained-based paradigms by designing an itemset pattern sampling approach enforcing a minimum frequency constraint.  {Diop~\cite{DBLP:conf/pakdd/Diop22} proposed a two-step procedure to sample high average utility itemsets under length constraints. Finally, Diop and Plantevit~\cite{DBLP:conf/bigdataconf/DiopP24} extended this work to sampling high-utility patterns from knowledge graphs by converting them into quantitative transactional databases.}

Richer patterns such as subgroups are also addressed by pattern sampling. In this context, Moens and Boley~\cite{DBLP:conf/ida/MoensB14} present a weighted controlled pattern sampling method for exceptional model mining. This approach divides the search space and applies targeted weighting, balancing efficiency and relevance to discover exceptional patterns.

To the best of our knowledge, there is only one  method for output space pattern sampling from numerical data~\cite{DBLP:conf/sdm/GiacomettiS18}. This method uses a multi-step approach. In contrast to \fips{} and \hips{}, it considers a continuous data space, samples according to a density measure, and requires the pattern size to be specified. {Our work falls into this class of approaches due to its ability to directly obtain the desired sampling distribution.}

\section{Sampling Interval Patterns Proportionally to their Frequencies}
\label{sec:contrib}

This section introduces \fips{}, an approach for sampling interval patterns proportionally to their frequency. The problem is formally defined as follows:

\[
Sampling_{k}(\mathcal{L}, \N, \freq) = \bigcup_{i=1}^{k} \{\IP_i \in \mathcal{L} \mid \IP_i \sim {\freq(\IP_i)}\}
\]

\subsection{\fips{} Key Ideas}
\label{sec:fips key ideas}

To draw a pattern proportionally to its frequency, it is necessary to determine the sum of the frequencies of all patterns in the solution space, i.e. $\sum_{\IP \in \mathcal{L}} \mathrm{freq}(\IP)$. Due to the size of the solution space $\lang$, it is intractable to compute this sum by enumerating each solution in the space. This limitation can be overcome by using a multi-step sampling procedure \cite{DBLP:conf/kdd/BoleyLPG11}. 
The key idea is to compute weights by grouping patterns according to their occurrences in the set of objects. Each group is then drawn in proportion to its calculated weight, then a pattern is selected uniformly within the chosen group.
Determining the number of patterns covering an object in binary data is straightforward {as it is simply equal to $2^{|g|}$}, where $|g|$ is the number of items present in the object $g$. In contrast, for numerical data, this task is more complex. {Unlike itemsets, which are contained within each transaction, interval patterns are not necessarily contained within the objects. Moreover, computing the exact number of interval patterns covering an object $g$ requires, for each attribute, considering all possible intervals that include the values appearing in $g$.} To address this challenge, we define a function $\NIP{}$ (Number of Interval Patterns) that computes the exact number of interval patterns covering an object.

%
Using this function, it is possible to calculate the sum of the frequencies of all patterns (the normalisation constant $Z$, see Section~\ref{sec:algo echantillonnage}) by considering only the patterns which cover an object. Thus, this method computes a probability distribution proportional to frequency without enumerating the entire solution space.


\subsection{Counting the Number of Interval Patterns Covering an Object}
\label{sec:NIP}

This section defines the function $\NIP{} : \ig \rightarrow \mathbb{N}$ which counts the exact number of interval patterns covering the object $\ig \in \G$. 
For a value $\val{}{} \in \N_\im$ appearing in an object $\ig \in \G$, we define:
\begin{itemize} 
    \item{ $\I{\val{}{}} = \{v \in \W_m \mid v \leq \val{}{}\}$ as the set of distinct values in attribute $\im$ that can be used as lower bounds in interval patterns covering the object $g$. }
    \item {$\J{\val{}{}} = \{v \in \W_m \mid v \geq \val{}{}\}$ as the set of distinct values in attribute $\im$ that can be used as upper bounds in interval patterns covering the object $g$.}
\end{itemize}

The  $\NIP(g)$ function is defined as follows: 

\begin{equation}
\label{eq:NIP}
\NIP(g) = \prod_{m \in \M} \quad {|\I{\val{}{}}| \cdot |\J{\val{}{}}|}
\end{equation}

{Equation~\ref{eq:NIP} relies on two terms. The first term $|\I{\val{}{}}|$ counts the number of distinct values in $\W_m$ that can serve as lower bounds for intervals associated to attribute $m$. Similarly, the second term $|\J{\val{}{}}|$ counts the number of distinct values in $\W_m$ that can serve as upper bounds for intervals associated to attribute $m$. The product $|\I{\val{}{}}| \cdot |\J{\val{}{}}|$ returns, for an object $g \in \G$ and an attribute $m \in \M$, the total number of intervals including the value $\val{}{}$. The exact number of interval patterns covering the object $\ig$ is given by the product of the possible intervals across all attributes in the database.}


\begin{example}
{
Consider the numerical database $\N$ presented in Table~\ref{dataset:nv:exemple}, the object $\ig_{3} \in \G$, and the attribute $m_{1} \in \M$. The distinct value of attribute $m_1$ in object $\ig_3$ is 3. The intervals associated to $m_1$ that include the value $3$ are the intervals  $[2,3]$, $[2,4]$, $[2,6]$, $[3,3]$, $[3,4]$ and $[3,6]$.
Thus, the total number of intervals including the value 3 for the attribute $m_{1}$ is 6. Referring to the terms in Equation~\ref{eq:NIP}, we have: $|\I{3}|\cdot |\J{3}| = |\{2,3 \}| \cdot |\{3, 4, 6\}| =6$. The total number of interval patterns covering $\ig_3$ is obtained by multiplying the number of intervals for each attribute (see Equation~\ref{eq:NIP}): $6 \times 4 \times 5 = 120$.
}
\end{example}

\subsection{\fips{} Sampling Algorithm}
\label{sec:algo echantillonnage}


This section describes the \fips{} sampling procedure (see Algorithm~\ref{algo:FIPS}). The algorithm starts by determining the number of interval patterns covering each object in the database $\N$ (line~\ref{algo:0}). {This is performed by} using the $\NIP{}$ function (see Section~\ref{sec:NIP}). Line~\ref{algoFips:Step1} implements the first step of the two-step sampling procedure: an object $g$ is sampled with a probability proportional to the number of interval patterns covering $g$. Thus, the draw is biased to select patterns covering a high number of objects. Then, in the second step (line~\ref{algo:2}), an interval pattern covering at least the object $g$ is built attribute by attribute through successive interval samplings. Specifically, for each attribute $m \in \M$, two values {$a_m \in \I{\val{}{}}$ and $b_m \in \J{\val{}{}}$} are uniformly sampled within each respective set. {These values form the interval associated to attribute $m$ and are concatenated into the pattern under construction (line~\ref{algo:concat}). Once all attributes have been associated to an interval, the resulting interval pattern $\IP$ is returned (line~\ref{algo:returnIP}).}  A sample of $k$ patterns is obtained by running $k$ times Algorithm~\ref{algo:FIPS}. 

\begin{algorithm}[h]
    {
    \textbf{Input :} A numerical database $\N$ \;
   \textbf{Output :} An interval pattern $\IP$ sampled proportionally to its frequency \;

\nonl~\\

       \textbf{Pre-processing:} Compute $w_{F}(g)= \NIP(g)$ for each object $g \in \G$\label{algo:0}\;

\nonl~\\
        
        \textbf{Step 1:} $g \sim  w_{F}(g)$ \quad \tcc*{Draw an object g proportionally to its weight $w_{F}$} \label{algoFips:Step1}

\nonl~\\
            \textbf{Step 2:} \label{algo:2}\
         \tcc*{Draw uniformly an interval pattern $\IP$ covering $g$.}
        
        $\IP \longleftarrow \langle \rangle$\;
        \ForEach{attribute $m \in \M $}{
        $a_m \longleftarrow$ Draw uniformly a value from {$\I{\val{}{}}$}\;
        $b_m \longleftarrow $  Draw uniformly a value from {$\J{\val{}{}}$}\;
        $\IP \longleftarrow \IP ++ [a_m,~ b_m] $  \tcc*{{Concatenate the interval $[a_m, b_m]$ to $\IP$ ($++$ denotes concatenation)}} \label{algo:concat}\
        }
            
\nonl~\\

            \textbf{Return } $\IP$ \label{algo:returnIP}\

    }
    \caption{{Sampling an interval pattern} proportionally to frequency (\fips{}) 
    \label{algo:FIPS}}
\end{algorithm}



\subsection{Sampling Distribution}
 In this section, we demonstrate that \fips{} samples interval patterns in proportion to their frequencies.

\vspace*{0.2cm}
\begin{property}
\label{prop:fips:distribution}
For a numerical database $\N$, Algorithm~\ref{algo:FIPS} samples an interval pattern $\IP$ proportionally to its frequency.
\end{property}

\begin{proof}
Let $\lang$ be the space of all interval patterns, and $Z = \sum_{\IP \in \lang } |\cover(\IP)|$ be the normalization constant, representing the sum of the frequencies of all interval patterns. Let $g^{*} \in \G$ be an object randomly drawn from step~1 of Algorithm~\ref{algo:FIPS}, and $\IP^{*}$ be an interval pattern sampled in step~2 of the same algorithm. Then:

\begin{align*}
P[\IP^{*}= \IP]&= \sum_{g ~\in~ \G} {P[\IP^{*}= \IP ~~\cap~~ g^{*}= g]}\\
&={\sum_{g ~\in~ \G} P[\IP^{*}= \IP ~|~ g^{*}= g] ~\cdot~P[g^{*}= g]}\\
&=  \sum_{g ~\in~ {\cover(\IP)}} \frac{1}{NIP(g)} \frac{NIP(g)}{Z}\\
&=  \sum_{\scriptsize g ~\in~{\cover(\IP)}} \frac{1}{Z} = \frac{|{\cover(\IP)}|}{Z} = \frac{{freq(\IP)}}{Z}
\end{align*}

where $Z= \sum_{g ~\in~ \G}~ NIP(g)$ (which is equal to $Z= \sum_{ \IP ~\in~ \mathcal{L}} ~|{\cover(\IP)}|  $)
\end{proof}




\subsection{Complexity Analysis of \fips{}}
In this section we conduct a computational complexity analysis for \fips{} that considers both data pre-processing and the sampling procedure. 

\vspace*{0.2cm}

\begin{property}
The overall time complexity of the \fips{} approach for sampling a single interval pattern is:

\[
O( |\G| \cdot |\M| + {log|\G| + |\M|})
\]

\end{property}

\begin{proof}
The numerical database $\mathcal{N}$ contains $|\G|$ objects and $|\M|$ attributes. In the worst case, each attribute has $|\G|$ distinct values. The time complexity of the pre-processing step, which uses the \NIP{} function, is $O(|\G| \cdot |\M|)$. This corresponds to computing the candidate sets $\mathcal{U}(\val{}{})$ and $\mathcal{A}(\val{}{})$ for each of the $|\M|$ attributes, where each computation takes $O(|\G|)$ time.

The time complexity of sampling an interval pattern using Algorithm~\ref{algo:FIPS} combines the cost of drawing an object in step~1 and sampling intervals for each attribute in step~2. {Drawing an object proportionally to its $\NIP(g)$ value is done by computing the cumulative sum of all $\NIP$ values, generating a random number scaled by the total sum, and locating the corresponding object via dichotomic search in $O(\log |\G|)$. Sampling intervals uniformly in step~2 takes $O(|\M|)$. Therefore, the overall asymptotic time complexity, including both the pre-processing and sampling phases, is $O(|\G| \cdot |\M| + \log |\G| + |\M|)$.}

\end{proof}

Sampling $k$ interval patterns requires $O(k(\log|\G| + |\M|))$ time in addition to the one-time pre-processing cost of $O(|\G| \cdot |\M|)$.

 \vspace*{0.2cm}

In an interactive pattern mining process, the analyst is often interested in patterns that cover a substantial region of the search space. For interval patterns, this corresponds to patterns with a large hyper-volume. However, considering only hyper-volume may lead to patterns covering only few observations. 
To avoid such undesirable patterns, we propose to consider both hyper-volume and frequency. The next section presents \hips{}, an extension of \fips{}, which samples patterns proportionally to the product of their frequency and hyper-volume.


\section{\hips{}: Incorporating Hyper-volume to \fips{}
} 	
\label{contrib:HFIPS}
	

This section presents \hips{}, a method that samples patterns proportionally to the product of their frequency and hyper-volume. The problem is formally defined as follows:

\[
Sampling_{k}(\mathcal{L}, \N, \freq \times \vol) = \bigcup_{i=1}^{k} \{\IP_i \in \mathcal{L} \mid \IP_i \sim {\freq(\IP_i) \cdot \vol(\IP_i)}\}
\]

\subsection{\hips{} Key Ideas}

\hips{} is built on the same core principles as \fips{} (cf. Section~\ref{sec:fips key ideas}). It computes, for each pattern in the solution space, the product of its hyper-volume and frequency without explicitly enumerating the entire space. This allows \hips{} to obtain the total sum $\sum_{\IP \in \mathcal{L}} (\freq(\IP) \cdot \vol(\IP))$ while avoiding a full enumeration. To achieve this, we introduce a new function, denoted $IPH$, which takes as input an object $g \in \G$ and computes the sum of the hyper-volumes of the patterns covering $g$. Summing the $IPH$ values across all objects leads to the normalization constant $Z$, as each pattern contributes proportionally to its frequency (see Section~\ref{sec:freqandvolume}).

\subsection{Efficiently Counting Total Hyper-volumes }
\label{sec:IH&TIA}

This section defines the function $IPH : \ig \rightarrow \mathbb{N}$, which computes the sum of the hyper-volumes of interval patterns covering the object $\ig \in \G$. $IPH$ is based on the idea of summing the lengths of all intervals that contain an object. We start by introducing this notion.

\vspace{0.5cm}
\textbf{Total Intervals Lengths.} To efficiently compute the hyper-volumes of interval patterns covering an object $\ig \in \G$, we first consider, for each attribute $m \in \M$, the sum of the lengths of all intervals containing the value $\val{}{} \in \N_\im$. To this end, we use the sets $\I{\val{}{}}$ and $\J{\val{}{}}$ introduced previously:

\begin{itemize} 
    \item $\I{\val{}{}} = \{v \in \W_m \mid v \leq \val{}{}\}$ as the set of distinct values in attribute $\im$ that can be used as lower bounds in interval patterns covering the object $g$. 
    \item $\J{\val{}{}} = \{v \in \W_m \mid v \geq \val{}{}\}$ as the set of distinct values in attribute $\im$ that can be used as upper bounds in interval patterns covering the object $g$.
\end{itemize}



The $TIL$ (Total Interval Lengths) function can be defined as follows:

\begin{equation}
\label{eq:TIL}
\TIL{m}{g} = \left(\sum_{x \in \J{\val{}{}}}  x \cdot |\I{\val{}{}} | \right) - \left(\sum_{x \in \I{\val{}{}}} x \cdot |\J{\val{}{}} |\right)
\end{equation}

where, the first term $(\sum_{x \in \J{\val{}{}}}  x \cdot |\I{\val{}{}} |)$ represents 
the sum of interval lengths by summing directly all the values of the upper bounds in $\J{\val{}{}}$ and 
multiplying this sum by the number of lower bounds in $\I{\val{}{}}$ since all upper bounds will be combined with each lower bound to form the intervals on the attribute including $\val{}{}$.
However, this term  alone overestimates the total length of all intervals, since it includes the upper bounds without subtracting the corresponding lower bounds that define the actual interval lengths.
On the opposite side, the second term $(\sum_{x \in \I{\val{}{}}} x \cdot |\J{\val{}{}} |)$, corrects this overestimation by subtracting
the sum of interval lengths by summing directly all the values of the lower bounds in $\I{\val{}{}}$ and 
multiplying this sum by the number of upper bounds in $\J{\val{}{}}$. Thus, ensuring with the first term that the length of each interval is computed exactly once. Consequently, the $TIL$ function leads to the exact sum of the lengths of all intervals containing $\val{}{}$ over the attribute $m$. 


\begin{example}
Consider the numerical dataset $\N$ shown in Table~\ref{dataset:nv:exemple}. For simplicity, we focus on attributes $m_1$ and $m_2$.  The Total Intervals Length for the object $g_1$ is computed for both attributes as follows:

\begin{itemize}
\item Total interval lengths for $m_1$: The value of $g_1$ for $m_1$ is $2$. $\TIL{m_1}{g_1}$  computes the total length of all intervals that contain the value $2$. 
Here, $\I{2} = \{2\}$ and $\J{2} = \{2,3,4,6\}$. The first part, $|\I{2}| \cdot \sum \J{2} = 1 \cdot (2+3+4+6)$, sums all possible upper bounds for intervals having $2$ as a lower bound, assuming they can all be paired with this single lower bound. However, this overestimates the total length, as it does not take into account the lower bound value itself. To correct this, the second term $|\J{2}| \cdot \sum \I{2} = 4 \cdot 2$ subtracts the repetitive contribution of the lower bound $2$, once for each possible upper bound. Therefore:
\begin{align*}
   \TIL{m_1}{g_1}  &= |\{2\}| \times (2+3+4+6) - |\{2,3,4,6\}| \times 2 \\
   &= (1 \times 15) - (4 \times 2) = 7
\end{align*}
This leads to the exact sum of lengths of the intervals $[2,2]$, $[2,3]$, $[2,4]$ and $[2,6]$, i.e., $0 + 1 + 2+ 4 = 7$.

\item Total interval lengths for $m_2$: The value of $g_1$ for $m_2$ is $8$. Similarly, $\TIL{m_2}{g_1}$  computes the total length of all intervals that contain the value $8$. Here, $\I{8} = \{7,8\}$ and $\J{8} = \{8,9,12\}$. The term $|\I{8}| \cdot \sum \J{8} = 2 \cdot (8+9+12)$ sums all upper bounds for each compatible lower bound. This overestimates the total interval lengths. The correction term $|\J{8}| \cdot \sum \I{8} = 3 \cdot (7+8)$ subtracts the repetitive contribution of the lower bounds. Therefore: 
\begin{align*}
    \TIL{m_2}{g_1} &= |\{7,8\}| \times (8+9+12) - |\{8,9,12\}| \times (7+8)\\
               &= (2 \times 29) - (3 \times 15) = 13
\end{align*}

The resulting value corresponds to the exact sum of lengths of the intervals $[7,8]$, $[7,9]$, $[7,12]$ $[8,8]$,$[8,9]$ and $[8,12]$ i.e., $1 + 2 + 5+ 0 + 1+ 4  = 13$.

\end{itemize}
\end{example}

\textbf{Total Hyper-volumes.} Once the total lengths of the intervals
have been computed for each attribute $m \in \M$,  the product of these interval lengths across all attributes gives the total hyper-volume of the interval patterns covering the object $g$. which is formally defined as follows: 

\begin{equation}
\label{eq:IPH}
    \IH{g} = \prod_{m \in \M} \TIL{m}{g}
\end{equation}
The proof of the $IPH$ function is provided in \ref{appendix:tilProof}


{

}

\begin{example}
Continuing from the previous example, the sum of the hyper-volumes of all interval patterns covering object $g_1$ is obtained by multiplying the total interval lengths for attributes $m_1$ and $m_2$:

\begin{align*}
    \IH{g_1} &= \TIL{m_1}{g_1} {~\cdot~}  \TIL{m_2}{g_1} \\
             &= 7 \times 13 = 91
\end{align*}
\end{example}

\subsection{\hips{} Sampling Algorithm}

This section describes the \hips{} sampling algorithm (see Algorithm~\ref{algo:hv_sampling}). The algorithm starts by computing, for each object $g$ in the database $\N$, the total hyper-volume of interval patterns covering $g$ (line~\ref{algohips:preprocessing}), using the \texttt{IPH} function (see Section~\ref{sec:IH&TIA}). Line~\ref{algoFips:Step1} implements the first step of the two-step sampling procedure, where an object $g$ is drawn with probability proportional to {the sum of hyper-volumes of the interval patterns that cover it}. In the second step (from line~\ref{algohips:step2}, to line \ref{algohips:concactinterval}), an interval pattern covering at least the selected object $g$ is sampled. The pattern is constructed attribute by attribute. For each attribute $m \in \M$, a lower bound $a_m$ is drawn from the set of candidate distinct values $\I{\val{}{}}$, using a weight function $w_{lb}$ that incorporates the cumulative contribution of potential upper bounds (line~\ref{algohips:step2lb}). This bias favors smaller values of $a_m$, which are more likely to lead to larger intervals. Then, an upper bound $b_m$ is sampled from the set of candidate values $\J{\val{}{}}$, using a weight function $w_{ub}$ that promotes wider intervals by taking into account the difference $b_m - a_m$ (line~\ref{algohips:step2ub}). Once both bounds are drawn, the corresponding interval is added to the {pattern under construction} (line~\ref{algohips:concactinterval}).

	\begin{algorithm}[h!]
		\caption{{Sampling an interval pattern proportionally to the product of its hyper-volume and frequency} (\hips{}) }
		\label{algo:hv_sampling}
	
		\textbf{Input:} A numerical database $\N $\;
		\textbf{Output:} An interval pattern $\IP$ sampled proportionally to its hyper-volume $\times$ frequency\label{algo1:preprocessing}\;
		\bigskip
		
		\textbf{Pre-processing:} Compute $w_{HF}(g)= \IH{g}$ for each object $g \in \G$ \label{algohips:preprocessing}\;
		\bigskip
		\textbf{Step 1:} Draw an object $g \in \G$ proportionally to it weight $w_{HF}$\label{algohips:step1} \;
		\bigskip
		$\IP \longleftarrow \langle\rangle $ \tcc*{ Interval pattern under construction} 
		\textbf{Step 2:} 
		\ForEach{attribute $m \in \M $}{ \label{algohips:step2}
		 \bigskip
			
			$\I{\val{}{}} = \{v \in \W_m \mid v \leq \val{}{}\}$  \tcc*{ Compute the candidate lower bound set}
			
			\bigskip
			
			$\J{\val{}{}} = \{v \in \W_m \mid v \geq \val{}{}\}$  \tcc*{ Compute the candidate upper bound set}
			\bigskip
			Draw a lower bound $a_m \in \I{\val{}{}}$ proportionally to $w_{lb}$, where: \label{algohips:step2lb}
			
			\begin{algomathdisplay}
w_{lb}(a_m) = \left(\sum_{x \in J_{g,m}} x\right) - |J_{g,m}| \cdot a_m
\end{algomathdisplay}
			
			Draw an upper bound $b_m \in \J{\val{}{}}$ proportionally to $w_{ub}$, where: \label{algohips:step2ub}
			\begin{algomathdisplay}
            w_{ub}(b_m) = b_m - a_m
                \end{algomathdisplay}
			$\IP \longleftarrow \IP ++ [a_m, b_m]$ \label{algohips:concactinterval}   \tcp*{ Adding $[a_m, b_m]$ to the interval pattern in construction}
		}
		
		\Return{$\IP$} 
	\end{algorithm}



\subsection{Sampling Distribution}
\label{sec:hips_Sampling Distribution}
 In this section, we demonstrate that \hips{} samples interval patterns in proportion to their frequencies $\times$ hyper-volumes.

\begin{property}
	For a numerical database $\N$, the \hips\ algorithm samples an interval pattern $\IP$ 
    with probability proportional to the product of  its hyper-volume and frequency. Formally:
		\[
		P[\IP] \propto \vol(\IP) \cdot \freq(\IP)
		\]
\end{property}

	\begin{proof}
		Let $\mathcal{L}$ be the set of all interval patterns. We define the normalization constant $Z$ as:
		\[
		Z = \sum_{\IP \in \mathcal{L}} \vol(\IP) \cdot \freq(\IP) = \sum_{g ~\in~ \G}~ \IH{g}.
		\]
		The \hips{} sampling procedure operates in two sequential steps.
		First, an object $g \in \G$ is sampled with a probability
			$P(g) = \frac{\IH{g}}{\sum_{g' \in \G} \IH{g'}}.$
		Second, given $g$ from the first step, an interval pattern $\IP$ that covers $g$ is drawn with a conditional probability proportional to its weight. 
		We demonstrate first that
			$P[\IP \mid g] = \frac{\vol(\IP)}{\IH{g}}.$
		Then, we demonstrate that the overall probability of drawing a particular interval pattern $\IP$  by \hips{} is equal to $P[\IP] = \frac{\vol(\IP)}{Z}  \times \freq(\IP)$.
			
		For a given object $g \in \G$ and for each attribute $m \in \M$, \hips{} constructs an interval $[\inff{w_m}, \supp{w_m}]$ by performing two sequential sampling steps. Consider the candidate lower and upper bound sets: $\I{\val{}{}} = \{v \in \W_m \mid v \leq \val{}{}\}  $ and $  \J{\val{}{}} = \{v \in \W_m \mid v \geq \val{}{}\}.$
		\begin{enumerate}
			\item \textbf{Lower Bound Selection:} 
			Each candidate lower bound $v \in \I{\val{}{}}$ is assigned a weight $w_{lb}(v)$. The probability of selecting a specific lower bound $\inff{w_m}$ is given by:
			
			    \begin{align*}
            P(\inff{w_m} \mid g) 
                &= \frac{w_{lb}(\inff{w_m})}{\sum_{v \in \I{\val{}{}}} w_{lb}(v)} 
                \quad \text{with} \\
                w_{lb}(v) 
            &= \left(\sum_{x \in \J{\val{}{}}} x\right) - |\J{\val{}{}}| \cdot v
                \end{align*}
			
	        The normalizing constant $\sum_{v \in \I{\val{}{}}} w_{lb}(v)$ corresponds to the Total Intervals Length (cf. Equation \ref{eq:TIL}). This allows the normalization constant to be computed directly as:
			
			\begin{align*}
			\sum_{v \in \I{\val{}{}}} w_{lb}(v) &= |\I{\val{}{}}| \cdot \left(\sum_{x \in \J{\val{}{}}} x\right)  - |\J{\val{}{}}| \cdot \left(\sum_{v \in \I{\val{}{}}} v\right)\\
			&= \TIL{m}{g}.
			\end{align*}
			
			Then:  \[P(\inff{w_m} \mid g) = \frac{w_{lb}(\inff{w_m})}{\TIL{m}{g}}\]
			
			\vspace*{0.2cm}
			
			\item \textbf{Upper Bound Selection:}  
	  	   Once the lower bound $\inff{w_m}$ has been sampled, the upper bound $\supp{w_m}$ is selected from the set $\J{\val{}{}}$ according to the following probability:
			\[
			P(\supp{w_m} \mid g, \inff{w_m}) = \frac{\supp{w_m} - \inff{w_m}}{w_{lb}(\inff{w_m})}.
			\]
			This formulation favors larger intervals, as the probability increases with the difference $\supp{w_m} - \inff{w_m}$.
		\end{enumerate}
		
Therefore, the overall conditional probability of selecting the interval $[\inff{w_m}, \supp{w_m}]$ for an attribute $m$ is given by:

\begin{align*}
P\bigl([\inff{w_m}, \supp{w_m}] \mid g\bigr) 
&= P(\inff{w_m} \mid g) \cdot P(\supp{w_m} \mid g, \inff{w_m}) \\
&= \frac{w_{lb}(\inff{w_m})}{\TIL{m}{g}} \cdot \frac{\supp{w_m} - \inff{w_m}}{w_{lb}(\inff{w_m})} \\
&= \frac{\supp{w_m} - \inff{w_m}}{\TIL{m}{g}}.
\end{align*}

Since the intervals for different attributes are selected independently, the probability of constructing the entire interval pattern $\IP = \langle [\inff{w_m}, \supp{w_m}] \rangle_{m \in \{1, \dots, |\M|\}}$ covering an object $g \in \G$ is:

\begin{align*}
P[\IP \mid g] &= \frac{\prod_{[\inff{w_m},\supp{w_m}] \in \IP} (\supp{w_m} - \inff{w_m})}{\prod_{m \in \M} \TIL{m}{g}}
\end{align*}

Using the definition of hyper-volume and Equation~\ref{eq:IPH}, we have:

\begin{align*}
P[\IP \mid g] &= \frac{\vol(\IP)}{\IH{g}}
\end{align*}

		By the law of total probability \cite{feller1991introduction},
		the overall probability of drawing a particular interval pattern $\IP$ is
		\[
		P[\IP] = \sum_{g \in \cover(\IP)} P[\IP \mid g] \cdot P(g).
		\]
		Where $P(g)$ is the probability of sampling $g$ according to Step 1 of Algorithm~\ref{algo:hv_sampling}.
		Substituting the expressions from the two steps results in:
		\[
		P[\IP] = \sum_{g \in \cover(\IP)} \frac{\vol(\IP)}{\IH{g}} \cdot \frac{\IH{g}}{\sum_{g' \in \G} \IH{g'}} 
		= \frac{\vol(\IP) \cdot \freq(\IP)}{\sum_{g' \in \G} \IH{g'}}.
		\]
		Defining $Z = \sum_{g' \in \G} \IH{g'}$, the normalization constant, completes the derivation:
		\[
		P[\IP] = \frac{\vol(\IP)}{Z}  \times \freq(\IP).
		\]
	\end{proof}

\subsection{Complexity analysis of \hips{}}
 
We now analyze the time complexity of the \hips{} approach. The analysis is structured into three stages: data pre-processing, object sampling and interval construction.

\begin{property}
The overall time complexity of \hips{} for sampling one interval pattern is:


\[
O\Big(|\G| \cdot |\M|+ log(|\G|)+  |\M|\cdot (|\G| + \log |\G|) \Big).
\]
\end{property}



\begin{proof}
The numerical database $\mathcal{N}$ contains $|\G|$ objects and $|\M|$ attributes. In the worst case, each attribute has $|\G|$ distinct values. For every object $g$ and attribute $m$, the candidate sets $\I{\val{}{}}$ and $\J{\val{}{}}$ are computed using a linear scan over $\W_m$, requiring $O(2 \cdot |\G|)$ time.

The overall sampling procedure can be divided into three phases:

\begin{itemize}
    \item \textbf{Pre-processing:} For each object $g \in \G$, we compute its weight using the function $\IH{g}$. This involves computing the function $\TIL{m}{g}$, for every attribute $m \in \M$, which takes $O(2 \cdot |\G|)$ time per attribute. Since $\IH{g}$ is the product over all attributes, the total pre-processing time is $O(|\M| \cdot 2 \cdot |\G|)$.

    \item \textbf{Step 1:} An object $g \in \G$ is drawn proportionally to its weight $\IH{g}$. This step is performed using a dichotomic search, which has a time complexity of $O(\log |\G|)$.

    \item \textbf{Step 2:} For each attribute $m \in \M$, the following substeps are performed:
    \begin{enumerate}
        \item For each lower bound candidate $a_m \in \I{\val{}{}}$, compute its weight using $w_{lb}$ in $O(|\G|)$ time.
        \item Draw a lower bound $a_m$ proportionally to $w_{lb}$ using a dichotomic search in $O(\log |\G|)$.
        \item For each upper bound candidate $b_m \in \J{\val{}{}}$, compute its weight using $w_{ub}$ in $O(|\G|)$ time.
        \item Draw an upper bound $b_m$ proportionally to $w_{ub}$ using a dichotomic search in $O(\log |\G|)$.
    \end{enumerate}
    Hence, Step 2 has an asymptotic total complexity of 
    \[
    O\big(|\M| \cdot (|\G| + \log |\G|)\big).
    \]
\end{itemize}

By combining all three phases, the asymptotic overall time complexity of the \hips{} sampling procedure is:

\[
O\Big(|\G| \cdot |\M|+ log(|\G|)+  |\M|\cdot (|\G|  + \log |\G|) \Big).
\]

\end{proof}

In practice, an analyst may wish to sample k patterns. Repeat \hips{} $k$ times requires  $O(k \cdot (log(|\G|) + |\M| \cdot (|\G| + log(|\G|))))$ in addition to the pre-processing time which is in O($|\M| \cdot |\G|$). However, this complexity can be reduced by storing the sets $\I{\val{}{}}$ and $\J{\val{}{}}$ in memory during pre-processing, which avoids recomputing them during sampling.
	

	
\subsection{{Sampling distribution and IPH}}
	
{The following property quantifies the sampling probability of an object to the hyper-volumes of its interval patterns. We show that an object $g$ is more likely to be sampled than under uniform sampling if and only if the sum of the hyper-volumes of the interval patterns that cover $g$ exceeds the average hyper-volume.}

	
	\begin{property}
		The sampling distribution satisfies:
		\[
		P(g) \geq \frac{1}{|\G|} \iff \IH{g} \geq \frac{1}{|\G|} \sum_{g' \in \G} \IH{g'}
		\]
	\end{property}

\begin{proof}
By construction, our algorithm selects an object $g \in \G$ with probability
\[
P(g) = \frac{\IH{g}}{\sum_{g' \in \G} \IH{g'}}.
\]
An object $g$ is therefore sampled with a probability at least as high as under uniform sampling (i.e., $\frac{1}{|\G|}$) if and only if its hyper-volume $\IH{g}$ is at least equal to the average hyper-volume:
\[
P(g) \geq \frac{1}{|\G|} \iff \IH{g} \geq \frac{1}{|\G|} \sum_{g' \in \G} \IH{g'}.
\]
Let $A = \sum_{g' \in \G} \IH{g'}$ denote the total hyper-volume. Then:
\[
P(g) = \frac{\IH{g}}{A}.
\]
The inequality $P(g) \geq \frac{1}{|\G|}$ is equivalent to:
\[
\frac{\IH{g}}{A} \geq \frac{1}{|\G|} \quad \Longleftrightarrow \quad \IH{g} \geq \frac{A}{|\G|}.
\]
Since $\frac{A}{|\G|}$ corresponds to the average hyper-volume, this shows that $g$ is more likely to be sampled than under uniform sampling if and only if its hyper-volume exceeds the average. 
\end{proof}


\section{Experiments and Results}
\label{sec:exp}

This section experimentally evaluates the performance of both \fips{} and \hips{} by addressing the following questions.

\begin{enumerate}
\item  How reliable is the frequency (resp. hyper-volume times frequency) of interval patterns sampled by \fips{} (resp. \hips{})?
\item  How does the long-tail phenomenon affect \fips{} and \hips{}?
\item  How diverse are the interval patterns sampled by \fips{} and \hips{}?
\item  How does \fips{} and \hips{} perform in term of CPU time?
\item  What is the relevance of the patterns sampled by \fips{} and \hips{} according to the plausibility criterion?
\end{enumerate}

\textbf{Selected databases.} Our experimental evaluation is conducted on a set of 13 numerical databases. {\it Glass}, {\it Iris}, {\it balance-Scale}, {\it Diabetes}, {\it sonar} and {\it heart} databases are from {the} UCI Machine Learning Repository\footnote{https://archive.ics.uci.edu/}, {while the 7 remaining are from the experimental protocol of our earlier work} \cite{DBLP:conf/cpaior/BekkouchaOBC24}. The number of numerical attributes, objects and distinct values for each database is presented in Table~\ref{table:benchmark}. The results of frequency, plausibility and CPU time experiments are presented for a reduced number of databases. Further results are available in \ref{appendix:expe} and at: { \footnotesize \url{https://github.com/djawed-bkh/Interval-Patterns-Sampling}}

\begin{table}[h]
    \centering
\scalebox{1}{
\begin{tabular}{|c||r|r|r|}\hline
    Datasets & $\# \M$ & $\# \G$ & $\#$ Distinct values \\\hline
    NT & 3 & 130 & 67 \\\hline
    AP & 5 & 135 & 674 \\\hline
    BK & 5 & 96 & 313 \\\hline
    Cancer & 9 & 116 & 900 \\\hline
    CH & 8 & 209 & 396 \\\hline
    Yacht & 7 & 308 & 322 \\\hline
    Iris & 4 & 150 & 123 \\\hline
    LW & 10 & 189 & 253 \\\hline
    Glass & 9 & 214 & 939 \\\hline
    balance-scale & 4 & 625 & 20 \\\hline
    diabetes & 8 & 768 & 1254 \\\hline
    heart & 13 & 270 & 384 \\\hline
    sonar & 60 & 208 & 11 256 \\\hline
\end{tabular}}
\caption{Datasets characteristics}
\label{table:benchmark}
\end{table}

\textbf{Compared approaches.} To the best of our knowledge, there is no approach for sampling interval patterns from numerical data, so we compare \fips{} and \hips{} with a first method that performs a uniform sampling of interval patterns. This first method uniformly draws an attribute $m \in \M$ and then uniformly draws the interval bounds from the values of the attribute $m$. This procedure is repeated until intervals are constructed for all the database attributes. However, as shown in Table~\ref{table:covVide}, this method samples many patterns with empty coverage, making them uninformative for an analyst.

To address this issue, we add a coverage control as described {in} Algorithm~\ref{algo:random} and we call \alea{} the resulting method. While there are available attributes in $\M^*$ (line \ref{uniform:while}), \alea{} uniformly selects and removes an attribute $\im$ from $\M^*$ (lines~\ref{uniform:selectM}-\ref{uniform:dropM}).
{Given the current coverage of the partially constructed interval pattern \( \IP \) (i.e. the subset of objects covered by the intervals already sampled in $\IP$), a candidate set $\N_{m}^{*}$ is defined (line~\ref{uniform:coverNm}).
The interval bounds $[a_m, b_m]$ for attribute $\im$ are then sampled by requiring that at least one bound appears in an object within the coverage of \( \IP \), thereby preventing empty coverage (line~\ref{uniform:bounds}).}
The new sampled interval is added to the partially constructed interval pattern \( \IP \) (line~\ref{uniform:pip}).
Finally, the sampled interval pattern is returned (line \ref{uniform:rip}). 
The source code of these methods and experimental results are available at  { \footnotesize \url{https://github.com/djawed-bkh/Interval-Patterns-Sampling}}


\begin{algorithm}[h]
    {
     \textbf{Input:} $\N$: Numerical database; \\
     \textbf{Output:} An interval pattern $\IP$ sampled uniformly with a non empty coverage \;
    $\IP \longleftarrow \langle \rangle$ \tcp*{ Interval pattern under construction}
    
    $\M^* \longleftarrow \{m_1,..., m_{|\M|} \}$\tcp*{Remaining attributes}
\While{$\M^* \neq \emptyset$}{ \label{uniform:while}

 $m \longleftarrow$ Draw uniformly an attribute from $\M^*$ \label{uniform:selectM}\;
 $\M^* \longleftarrow \M^* \backslash {\{m\}} $ \label{uniform:dropM}\;
$\N_{m}^{*} \longleftarrow \{v_{g,m} \quad |\quad g \in cover(\IP) \}$
\tcp*{ Values from $\N_m$ appearing in objects of the current coverage of $\IP$} \label{uniform:coverNm}
        $[a_m, b_m] \longleftarrow$ Draw uniformly values $a_m, b_m$  such that: $$a_m \leq b_m \text{ and } \left\{\begin{array}{lll}
   a_m \in  \N_m \land b_m \in \N_{m}^{*} &\\ 
 or &\\
  a_m \in  \N_{m}^{*} \land b_m \in \N_{m} &
\end{array}\right.$$
\tcc*{ 
selecting from $\N_m^*$ ensures the coverage of at least one object}  \label{uniform:bounds}

$\IP \longleftarrow \IP ++ [a_m, b_m]$ \label{uniform:pip}\;
}     
            \Return $\IP$ \label{uniform:rip}\;

    }

\caption{\alea{}: sampling of interval patterns \label{algo:random}}
\end{algorithm}

\begin{table}[h]
    \centering
\scalebox{1}{
\begin{tabular}{|c||c|}\hline
    Databases & Empty coverage (\%) \\\hline
    NT & 8 \\\hline
    AP & 76 \\\hline
    BK & 66 \\\hline
    Cancer & 98 \\\hline
    CH & 96 \\\hline
    Yacht & 81 \\\hline
    Iris & 56 \\\hline
    LW & 86 \\\hline
    Glass & 99 \\\hline
    balance-scale & 0 \\\hline
    diabetes & 92 \\\hline
    heart & 96 \\\hline
    sonar & 100 \\\hline
\end{tabular}}
\caption{Percentage of patterns having an empty coverage when sampling interval patterns according to a uniform distribution without coverage control (10 000 sampled patterns).}
\label{table:covVide}
\end{table}

\subsection{Distribution of Sampled Patterns and Long Tail Phenomenon}
\label{sec:freqandvolume}

Sections~\ref{sec:xp_freq} and \ref{sec:xp_hypervol} address Question 1. We evaluate the patterns sampled by \fips{} and \hips{} according to their respective interestingness measures. These patterns are compared to  patterns sampled uniformly using \alea{}. We address Question 2 in Section~\ref{sec:longtaileval} by evaluating the robustness of each approach with regard to the long tail phenomenon.

\subsubsection{Evaluation of Patterns {with Respect to} the Frequency}
\label{sec:xp_freq}

\begin{figure}[h]
    \begin{minipage}[b]{0.45\linewidth}
        \centering
        \includegraphics[width=\linewidth]{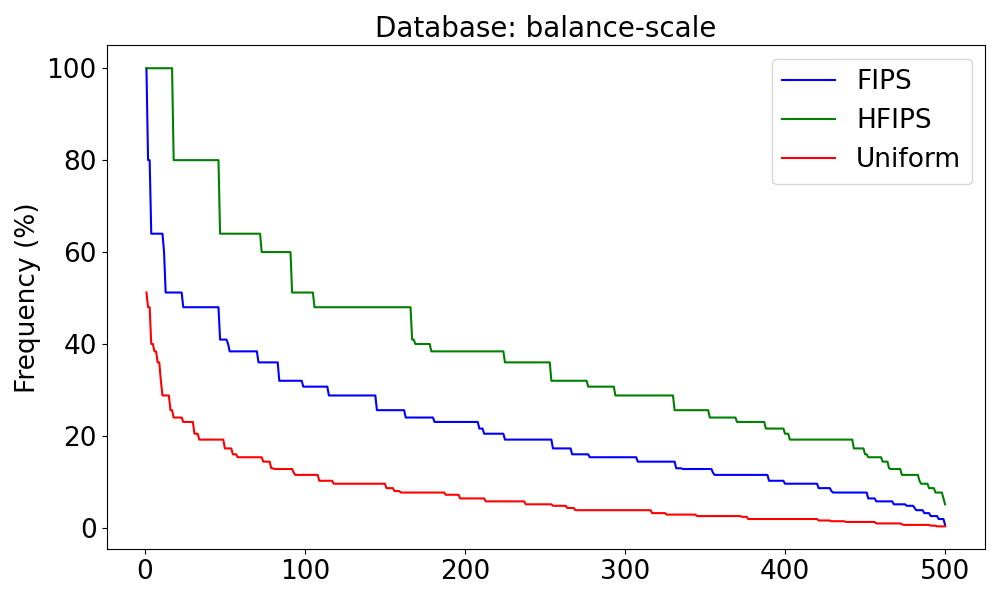}
    \end{minipage}
    \hspace{0.05\linewidth}
    \begin{minipage}[b]{0.45\linewidth}
        \centering
        \includegraphics[width=\linewidth]{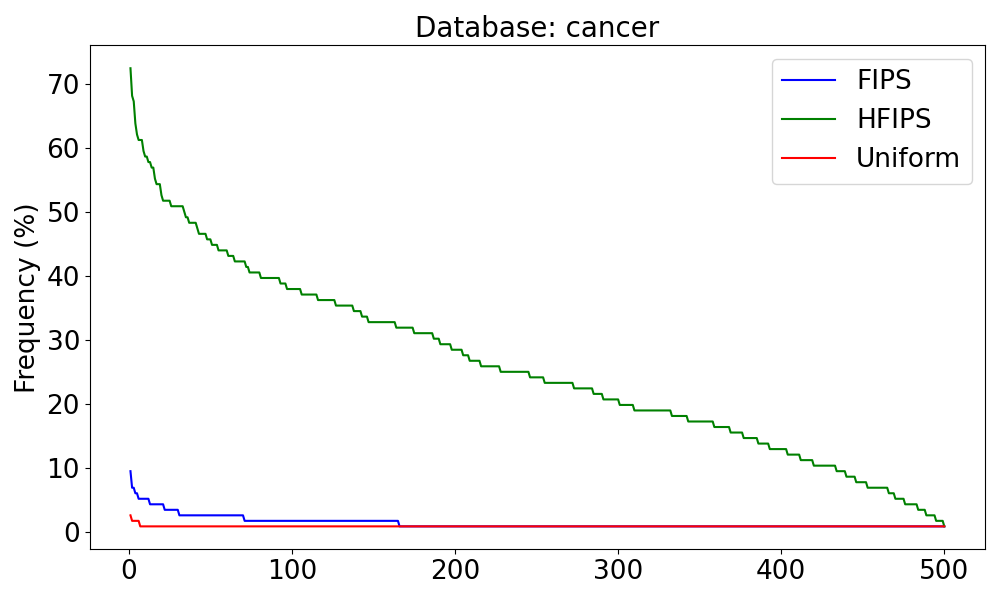}
    \end{minipage}

     \begin{minipage}[b]{0.45\linewidth}
        \centering
        \includegraphics[width=\linewidth]{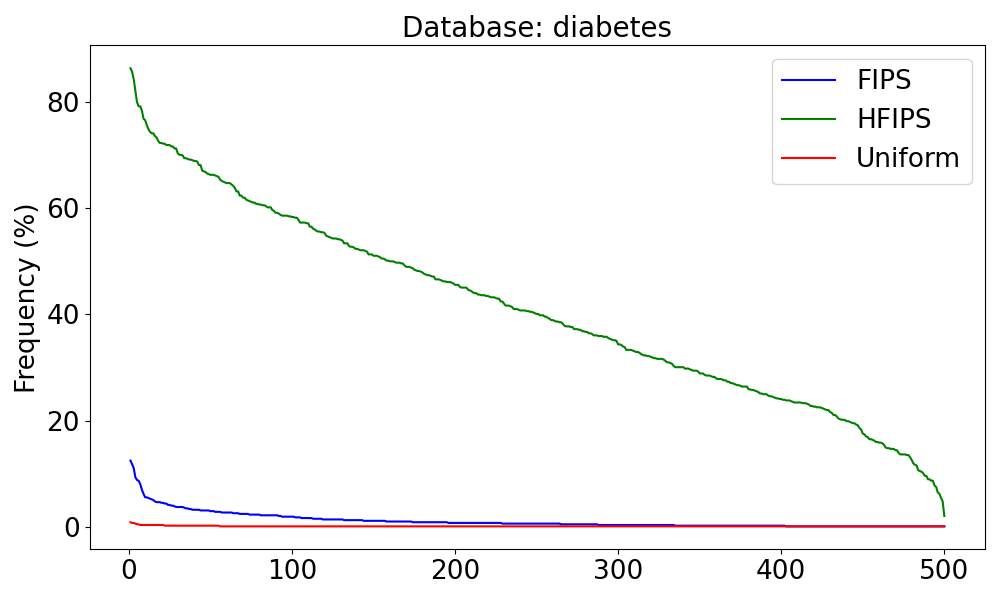}
    \end{minipage}
    \hspace{0.05\linewidth}
    \begin{minipage}[b]{0.45\linewidth}
        \centering
        \includegraphics[width=\linewidth]{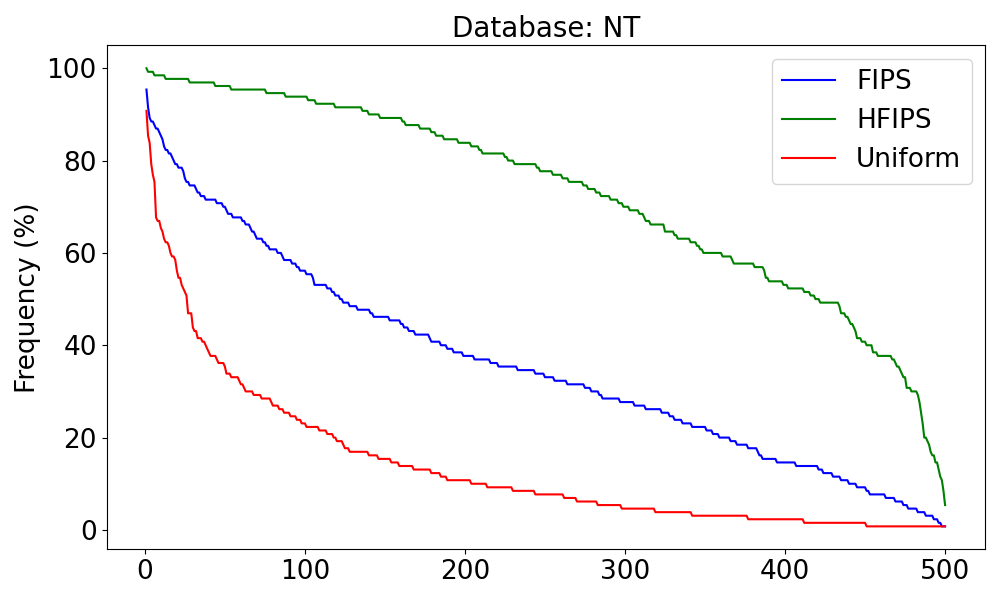}
    \end{minipage}
    \caption{ Frequency evaluation for 500 patterns sampled with \fips{} and the \alea{} methods }
    \label{exp:frequence}
\end{figure}

Figure~\ref{exp:frequence} presents the frequencies of 500 patterns sampled by \fips{}, \hips{} and \alea{}. The patterns are sorted in descending order of frequency, and the results are averaged over 10 iterations. 

We observe in all databases that \fips{} samples patterns with higher frequencies than those obtained with \alea{}. This can be attributed to Step~1 of \fips{}, which is biased towards selecting objects covered by a large number of patterns, leading to higher frequency. {\hips{} amplifies this frequency bias by jointly considering hyper-volume with frequency into the sampling procedure}.



\subsubsection{Evaluation of Patterns {with Respect to} Hyper-volume  $\times$ Frequency}
\label{sec:xp_hypervol}
\begin{figure}[h]
    \begin{minipage}[b]{0.45\linewidth}
        \centering
        \includegraphics[width=\linewidth]{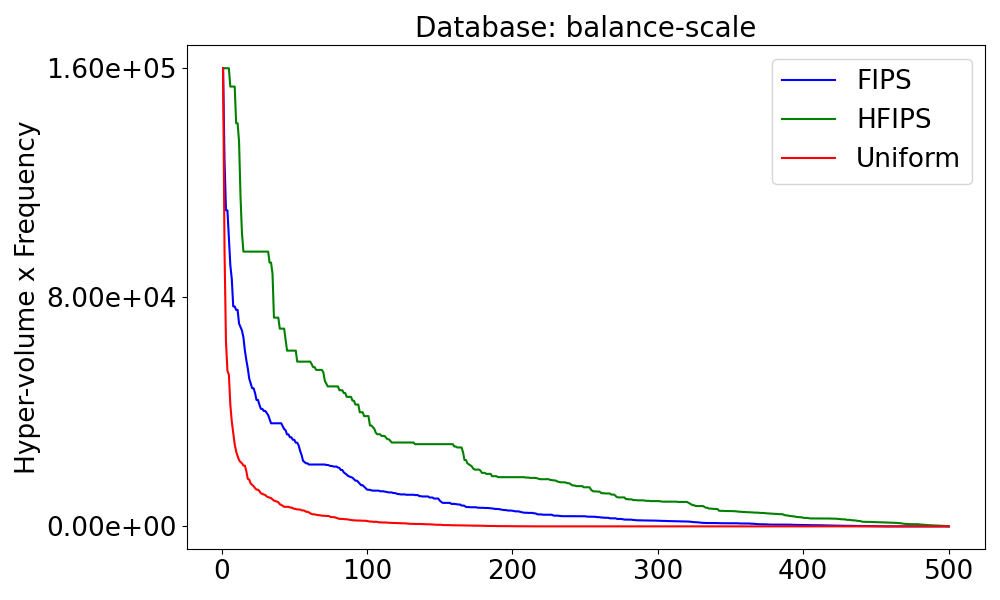}
    \end{minipage}
    \hspace{0.05\linewidth}
    \begin{minipage}[b]{0.45\linewidth}
        \centering
        \includegraphics[width=\linewidth]{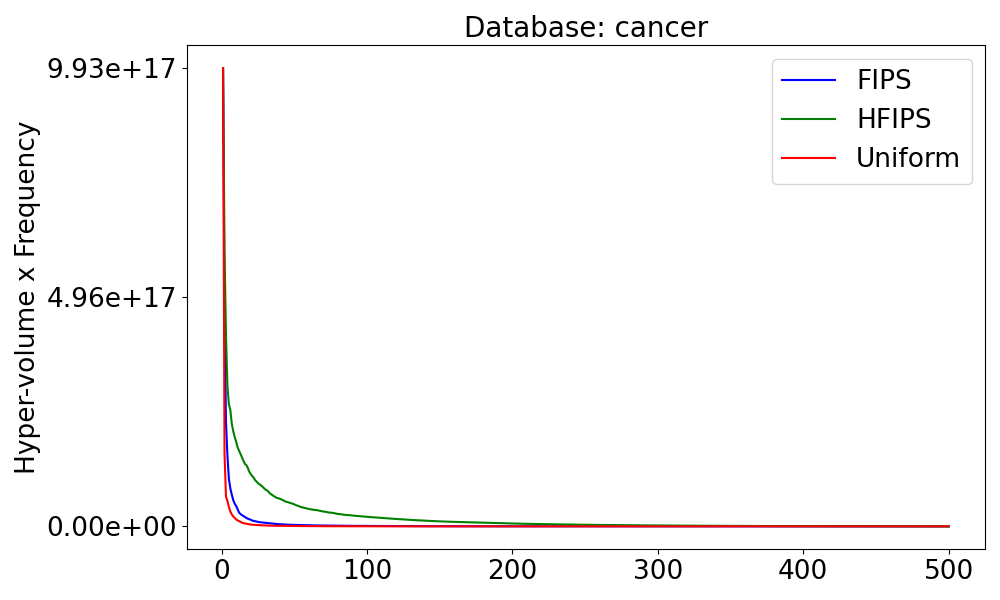}
    \end{minipage}

     \begin{minipage}[b]{0.45\linewidth}
        \centering
        \includegraphics[width=\linewidth]{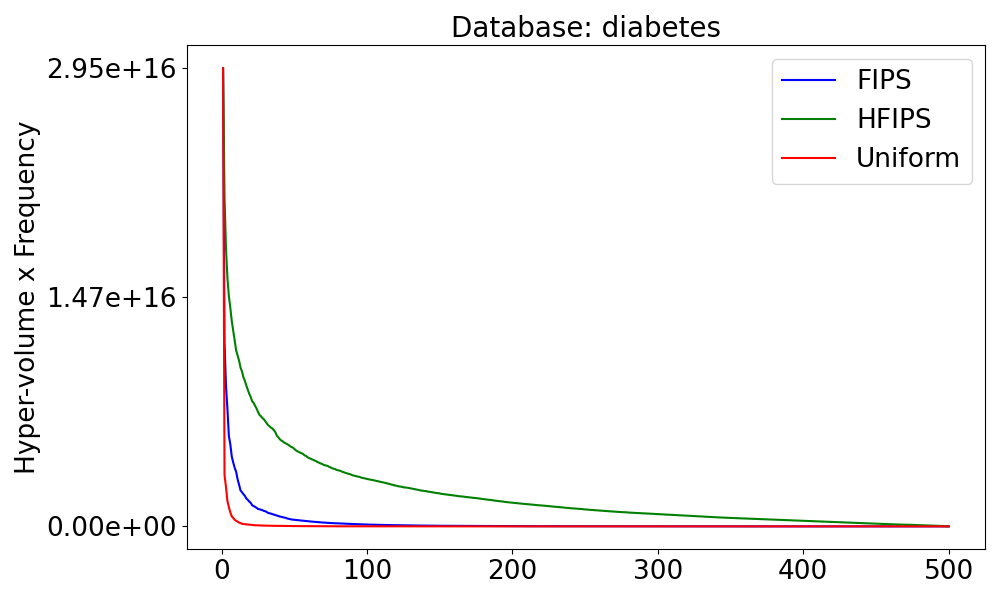}
    \end{minipage}
    \hspace{0.05\linewidth}
    \begin{minipage}[b]{0.45\linewidth}
        \centering
        \includegraphics[width=\linewidth]{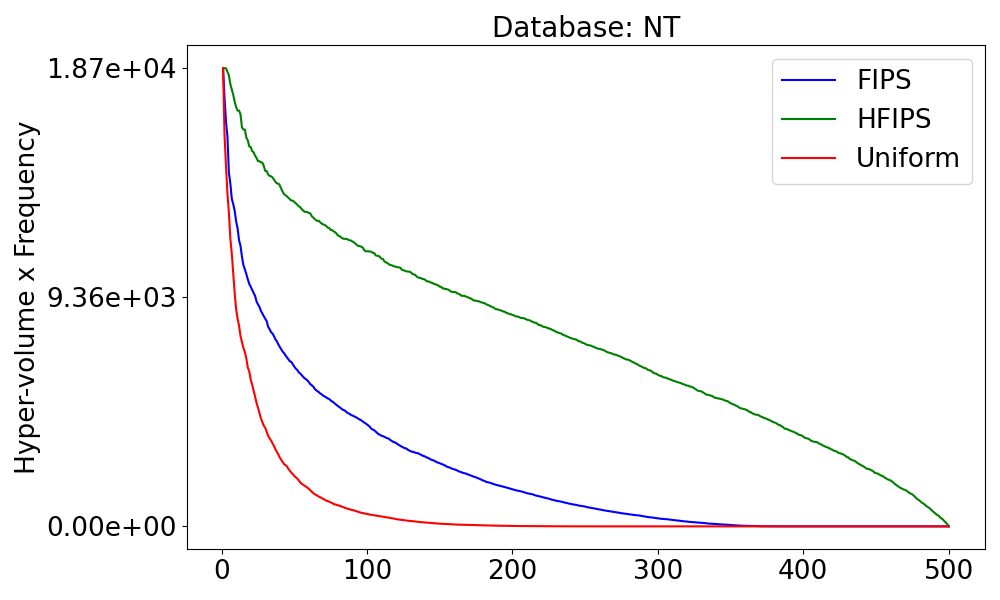}
    \end{minipage}
    \caption{ Volume evaluation for 500 patterns sampled with \hips{} and the \alea{} methods }
    \label{exp:volume}
\end{figure}

Figure \ref{exp:volume} shows the values of the product of hyper-volume and frequency of 500 patterns sampled by \fips{}, \hips{} and \alea{}, with patterns sorted in descending order according to this measure. The results are averaged over 10 iterations. 

For all databases, \fips{} samples patterns with higher values 
compared to those obtained with \alea{}. This is explained by the fact that \fips{} favors frequent patterns. {Since frequent interval patterns cover more database observations}, they are also more likely to have wider intervals, especially when the observations are scattered, resulting in higher volumes and consequently higher values of frequency $\times$ hyper-volume.
For all databases, \hips{} samples patterns with higher values compared to those obtained with \fips{}. 
This difference is due to  the sampling procedure of \hips{}, which in {Step 1 biases} the object selection towards those covered by patterns with both high frequencies and hyper-volumes and, in Step 2, biases the selection of interval bounds towards wide intervals. Resulting patterns have high frequencies with large hyper-volumes.

\subsubsection{Long Tail Phenomenon}
\label{sec:longtaileval}
Question~2 related to the long tail phenomenon is addressed in this section. The \textit{long tail phenomenon} is a statistical effect introduced { by Bryson~\cite{Bryson01021974} and popularized in the context of business by Anderson~\cite{anderson2006longtail}. It affects sampling techniques by leading to the sampling of uninteresting patterns.}
The long tail is characterized by an unbalanced distribution of sampled patterns with respect to the considered measure. A small number of patterns with high measure values form the {``head''}, while a large number of patterns with low measure values form the {``tail''}. In the following, we evaluate the robustness of \fips{} and \hips{} to the long tail phenomenon, each one according to its respective interestingness measure, namely frequency for \fips{} and the product of frequency and hyper-volume for \hips{}.

\vspace*{0.2cm}

\textbf{Robustness of \fips{} to the long tail phenomenon. }
Figure \ref{exp:frequence} shows that \fips{} is significantly less affected by the long-tail phenomenon than \alea{}. For the {\it diabetes} and {\it cancer} databases, 65\% and 68.2\% of patterns generated by \fips{} occur in the tail (frequency below 1\%),  in comparison to approximately 99\% with \alea{}. For the {\it balance-scale} and {\it NT} databases, \fips{} produces a negligible number of low-frequency patterns, making the tail almost undetectable. In contrast, \alea{} produces 9.4\% and 10\% of patterns with frequencies below 1\% for the balance-scale and NT databases, respectively.

\vspace*{0.2cm}

\textbf{Robustness of \hips{} to the long tail phenomenon} Similarly to \fips{}, Figure~\ref{exp:volume} shows that \hips{} is significantly less affected by the long-tail phenomenon than \alea{}, with respect to the considered interestingness measure. In the \textit{NT} database, the tail is almost undetectable. In the \textit{balance-scale}, \textit{cancer}, and \textit{diabetes} datasets, we observe that \hips{} consistently delays the appearance of the tail. For instance, in \textit{balance-scale}, only 50\% of the sampled patterns fall within the tail, compared to 86\% for \alea{}. In \textit{cancer}, the difference is even more pronounced: 85\% for \hips{} versus 99\% for \alea{}. Similarly, in \textit{diabetes}, 60\% of the \hips{} patterns appear in the tail, compared to 99\% for \alea{}. 

\subsection{Diversity Evaluation}
Sections~\ref{sec:eqclassdiversity} and \ref{sec:jaccardDIversity} address Question 3 on the diversity of sampled patterns. Sampling diverse patterns is essential in an interactive mining process, as it allows the analyst to explore multiple areas of the solution space.  In this section, we use various metrics to assess the diversity of patterns sampled by \fips{}, \hips{} and \alea{}. Section~\ref{sec:eqclassdiversity} deals with the diversity of equivalence classes among the sampled patterns, while Section~\ref{sec:jaccardDIversity} focuses on their overlap in terms of covered objects.

\subsubsection{Equivalence Classes Diversity}
\label{sec:eqclassdiversity}
\begin{figure}[h]
    \centering
    \includegraphics[width=1\linewidth]{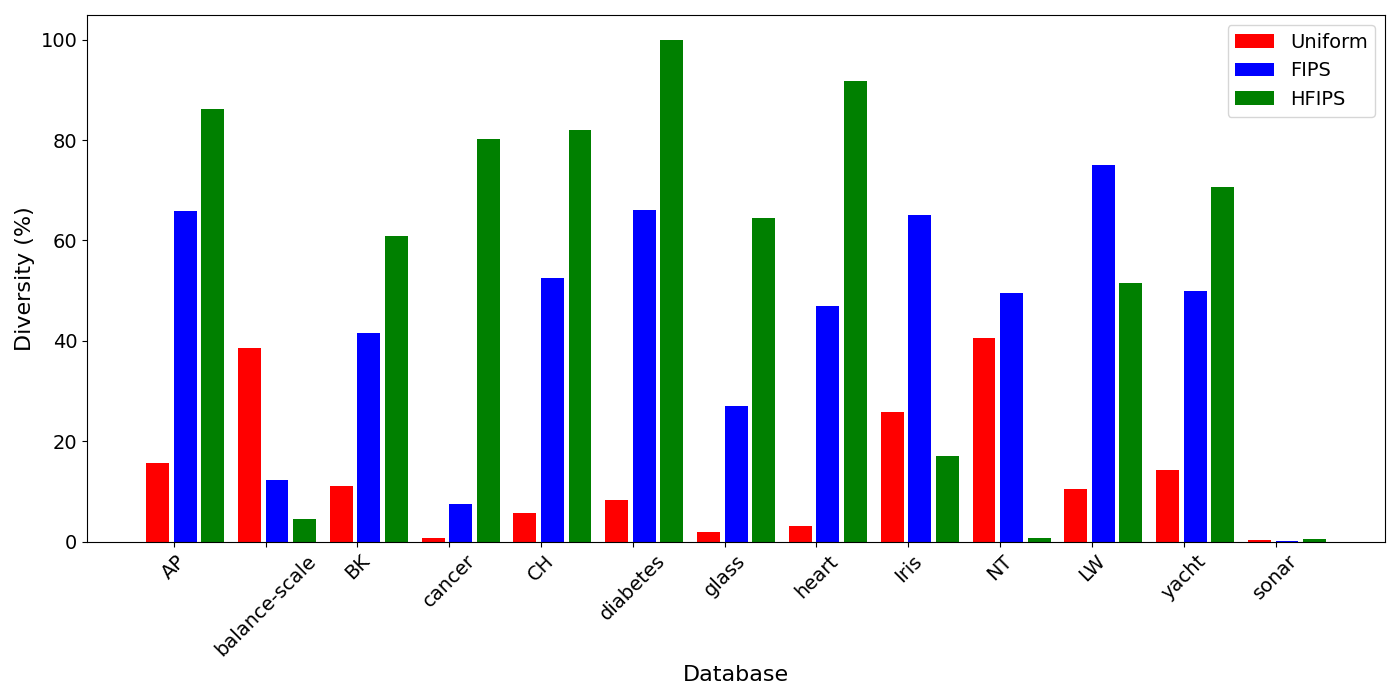}
    \caption{Diversity evaluation of \fips{}, \hips ~and \alea{} methods}
    \label{exp:DiversitySDM}
\end{figure}

To evaluate the diversity of \fips{}, \hips{} and \alea{} in terms of equivalence classes, we use the measure presented in \cite{DBLP:conf/sdm/GiacomettiS18}, which is defined as follows:  $$ diversity(K, \N) = \frac{|\{cover(\IP_1, \N), \dots, cover(\IP_{|K|}, \N)\}|}{|K|}$$ where $|K|$ is the number of sampled patterns. Figure~\ref{exp:DiversitySDM} shows that patterns sampled with \hips{} and \fips{} have a greater diversity in equivalence classes than those sampled with \alea{}, except in the case of the {\it balance-scale} database. This can be attributed to the higher frequency of patterns produced by \hips{} and \fips{} (see Section~\ref{sec:freqandvolume}), which results in wider coverage and, consequently, greater variation in the covered objects across samples,  thereby increasing diversity in terms of equivalence classes. The exception observed for the {\it balance-scale} dataset can be explained by its specific characteristics, in particular the limited number of distinct values and numerical attributes (see Table~\ref{table:benchmark}).  Since \fips{} and \hips{} are biased towards high frequency and high hyper-volume patterns, they tend to sample the same objects repeatedly, resulting in similar intervals and reduced diversity.
The higher diversity of \hips{} in comparison to \fips{} can be explained by the higher frequencies of patterns sampled by \hips{}, resulting in greater variation in the coverages. 


\subsubsection{Coverage Overlapping}
\label{sec:jaccardDIversity}
Since the metric presented in Section \ref{sec:eqclassdiversity} does not consider the overlap of coverages between the sampled patterns, we complement the diversity evaluation 
with a Jaccard index-based measure. We evaluate the coverage diversity of 500 patterns obtained with \fips{}, \hips{} and \alea{}, using the Jaccard index. As this metric compares pairs of patterns, it cannot be applied directly to a large set. Therefore, we use cumulative distribution functions (CDF) to visually represent the distribution of Jaccard indices computed for all pairs formed by our 500 patterns. Consider a set of \( k=500 \) interval patterns \( \boldsymbol{\IP}=\{\IP_1, \dots, \IP_{500}\} \) and a numerical database \( \N \).
\[
CDF(\N, \boldsymbol{\IP}, \theta) = \frac{ 2  \times \left| \{(i,j) : Jac(\IP_i, \IP_j) \leq \theta ,~~ 1 \leq i < j \leq k,~~ \IP_i, \IP_j \in \boldsymbol{\IP}  \} \right|}{k(k-1)}
\]

\begin{figure}[h]
    \begin{minipage}[b]{0.45\linewidth}
        \centering
        \includegraphics[width=\linewidth]{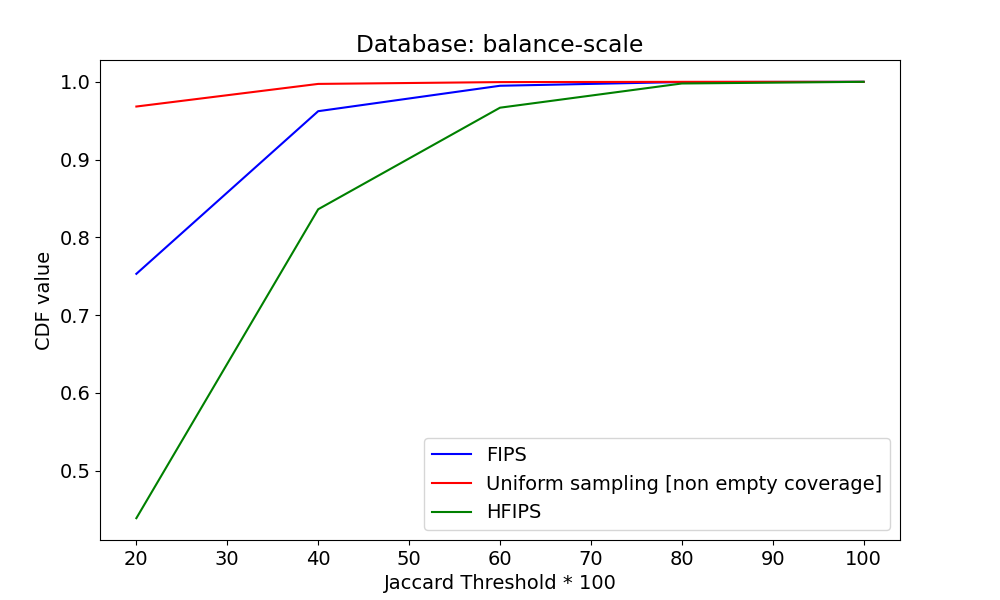}
    \end{minipage}
    \hspace{0.05\linewidth}
    \begin{minipage}[b]{0.45\linewidth}
        \centering
        \includegraphics[width=\linewidth]{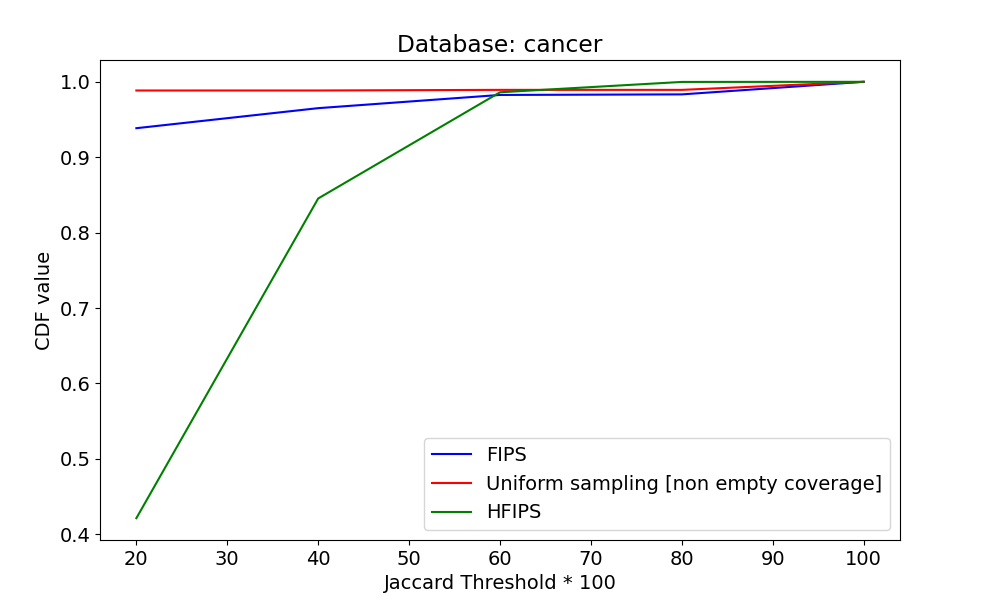}
    \end{minipage}

     \begin{minipage}[b]{0.45\linewidth}
        \centering
        \includegraphics[width=\linewidth]{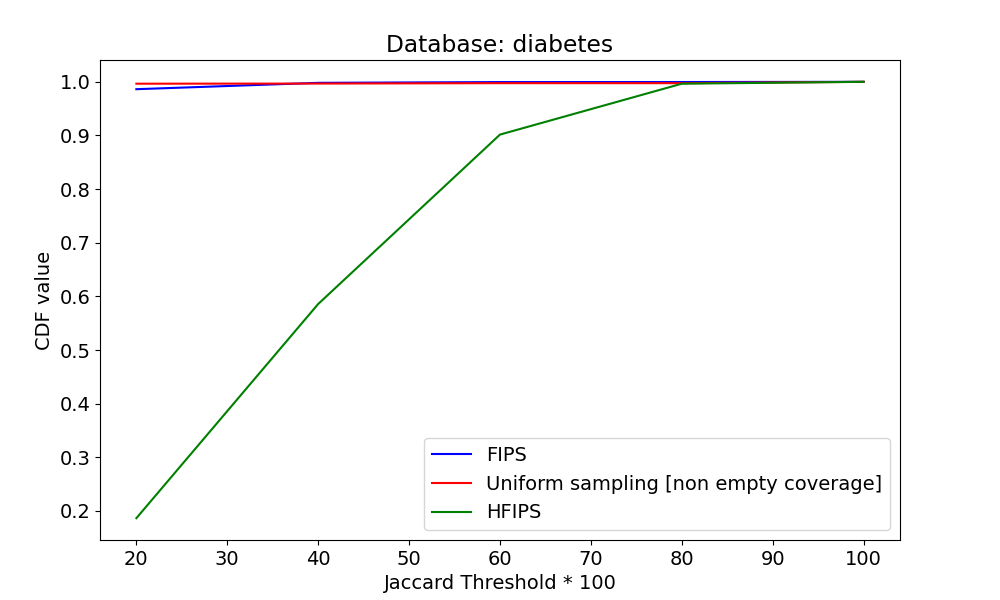}
    \end{minipage}
    \hspace{0.05\linewidth}
    \begin{minipage}[b]{0.45\linewidth}
        \centering
        \includegraphics[width=\linewidth]{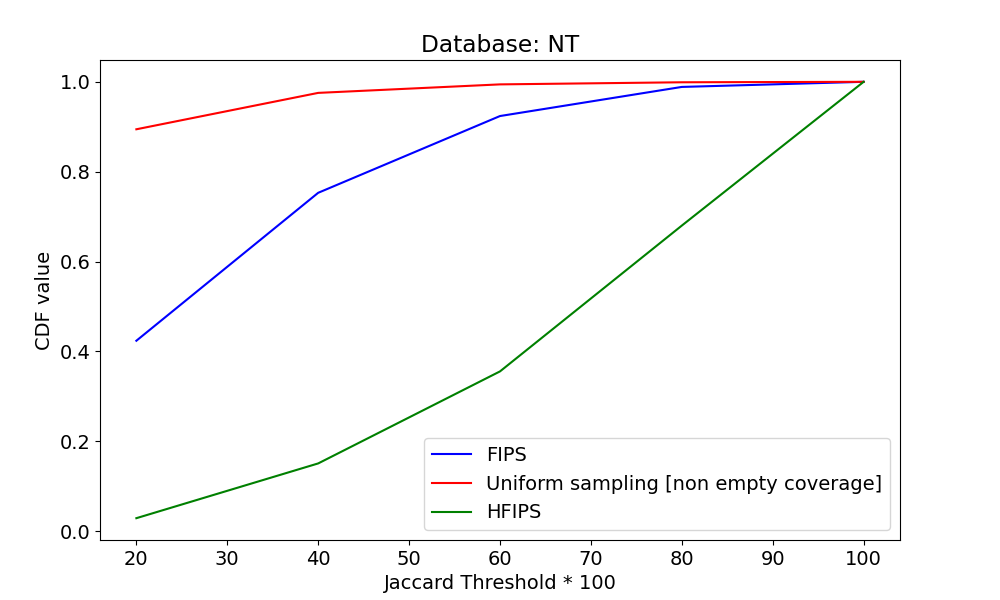}
    \end{minipage}
    \caption{ Coverage diversity evaluation for 500 patterns sampled with \fips{}, \hips ~and the \alea{} method }
    \label{exp:jaccardDiversity}
\end{figure}

Figure~\ref{exp:jaccardDiversity} shows that, for all databases, the cumulative distribution function curve of \hips{} has a lower area compared to \alea{}. This indicates that \hips{} generates patterns with higher similarity in terms of coverage than \alea{}. For small Jaccard indices, \hips{} returns a small number of patterns with low coverage similarity. However, as the similarity threshold increases, the number of patterns produced by \hips{} also grows, meaning that this method generates patterns that share the same objects in their coverage. In contrast, \alea{} shows greater coverage diversity: a large number of patterns have low Jaccard indices, and the curve generally remains linear as the similarity threshold increases. This means that there is no significant addition of highly similar patterns. This observation can be explained by the fact that \alea{} tends to sample patterns covering a small number of objects, reducing the probability that these patterns share objects with others from the same approach. In contrast, \hips{}, which samples patterns covering a large number of objects, increases the probability of sampling other patterns that share a subset of objects.

For the \fips{} method, we observe that for the \textit{NT} and \textit{balance-scale} databases, the cumulative distribution function curve has a lower area compared to the random sampling approach. This indicates that \fips{} generates patterns with higher similarity in terms of coverage than \alea{}. This can be explained by two factors: first, \fips{} samples patterns with greater frequencies than \alea{} (see Figure \ref{sec:freqandvolume}), meaning that patterns obtained with \fips{} are more likely to share the same objects in their coverages, leading to lower diversity. The second factor concerns the database characteristics (see Table \ref{table:benchmark}). We observe that \textit{NT} and \textit{balance-scale} have a small number of distinct values, which leads \fips{} to often sample the same patterns.
Finally, for the \textit{diabetes} and \textit{cancer} databases, we observe that the diversity of \fips{} is almost as high as that of \alea{}. This can be explained by the large number of distinct values in these databases, which leads \fips{} to sample more diversified patterns.

\subsection{CPU Time Evaluation}

\begin{figure}[h]
    \begin{minipage}[b]{0.45\linewidth}
        \centering
        \includegraphics[width=\linewidth]{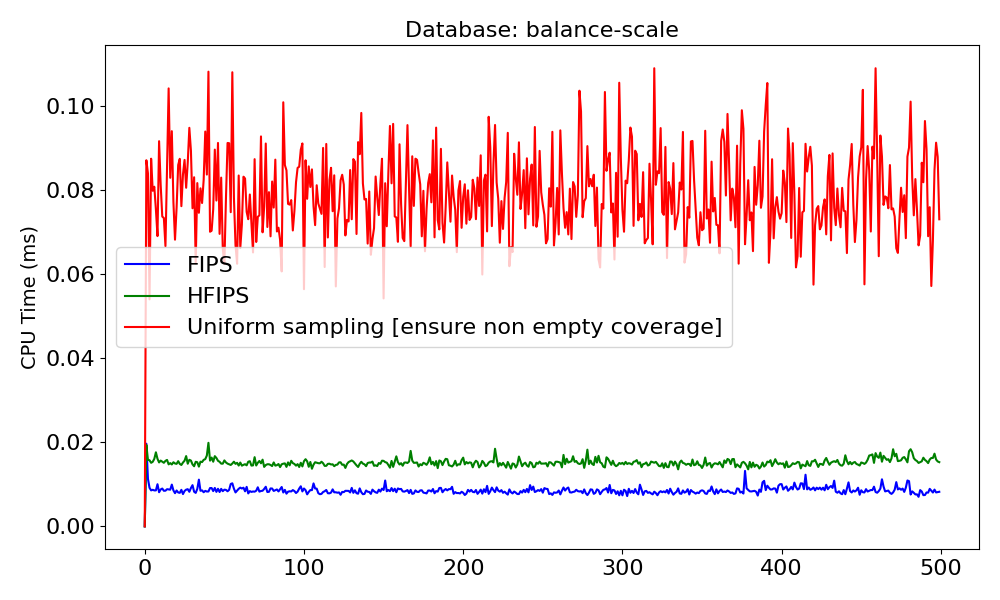}
    \end{minipage}
    \hspace{0.05\linewidth}
    \begin{minipage}[b]{0.45\linewidth}
        \centering
        \includegraphics[width=\linewidth]{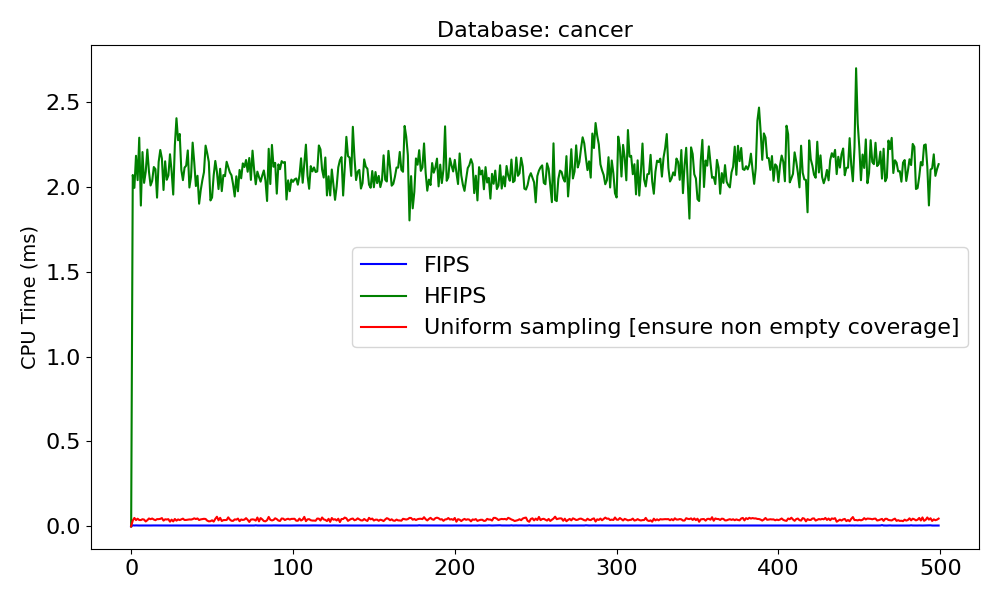}
    \end{minipage}

     \begin{minipage}[b]{0.45\linewidth}
        \centering
        \includegraphics[width=\linewidth]{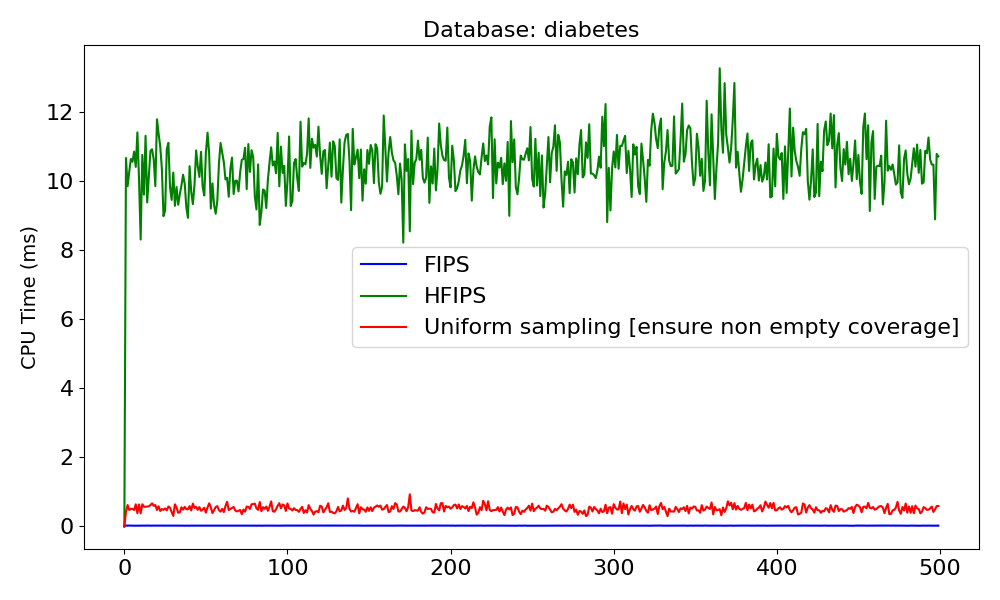}
    \end{minipage}
    \hspace{0.05\linewidth}
    \begin{minipage}[b]{0.45\linewidth}
        \centering
        \includegraphics[width=\linewidth]{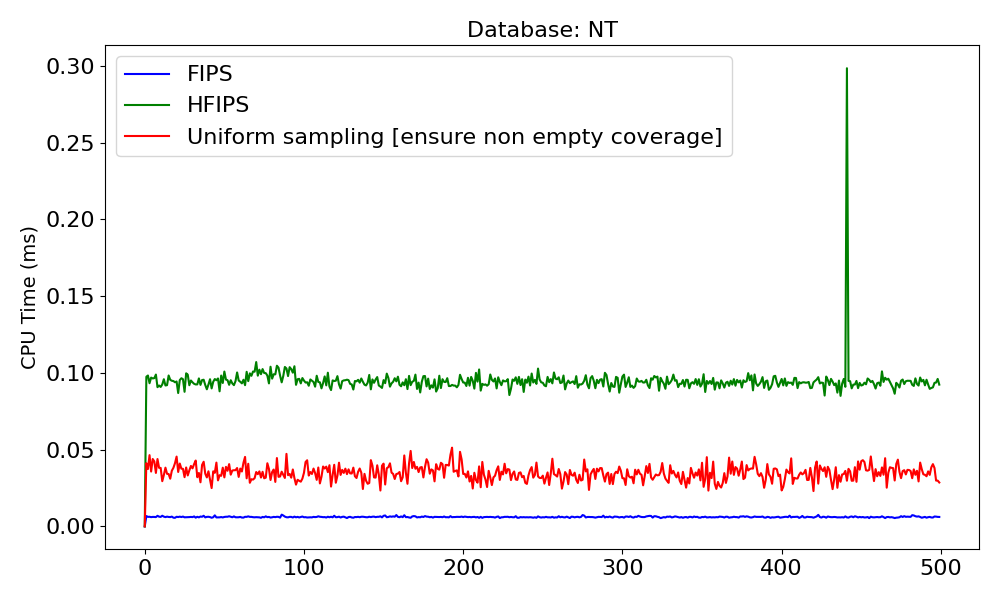}
    \end{minipage}
    \caption{
    CPU time evolution for a set of 500 patterns sampled by the \fips{}, \hips{} and \alea{} methods }
    \label{exp:tempsCPU}
\end{figure}

This section deals with running time (Question 4). Figure~\ref{exp:tempsCPU} presents the CPU time\footnote{Configuration: Intel Core i7, 11th generation, 3 GHz, 8 cores, 30 GB RAM} required to sample each of 500 interval patterns using \fips{}, \hips{}, and \alea{}. The results are averaged over 10 iterations.

We observe that for most of the databases, \fips{} is faster than both \alea{} and \hips{}. This can be explained by the fact that \alea{} requires, for each sampled interval in the pattern, a search for distinct values from the current coverage. This is achieved by recalculating the current coverage of the pattern being constructed for each newly sampled interval ($|\M|$ times in total). This procedure also explains the oscillations observed for this method. If the gap between the bounds of the first sampled intervals is large, the coverage computation time will be high. In contrast, if the gap is small, the computation time will be lower. With the exception of \textit{balance-scale}, this time is longer for \hips{} than for \alea{} and \fips{}. This is due to the second step of its algorithm (see Algorithm~\ref{algo:hv_sampling}), where the weight computation of the values that are candidates for being bound takes more CPU time. The exception on \textit{balance-scale} can be explained by the small number of distinct values present in this database (see Table~\ref{table:benchmark}), which results in a smaller number of candidate values for being bounds, thus requiring less computation time for the second step of \hips{} (see Algorithm~\ref{algo:hv_sampling}). Note that in {all cases}, run-times remain very short as the time scale is measured in milliseconds.

\subsection{Plausibility Evaluation}

\begin{figure}[h]
    \centering
    \includegraphics[width=1\linewidth]{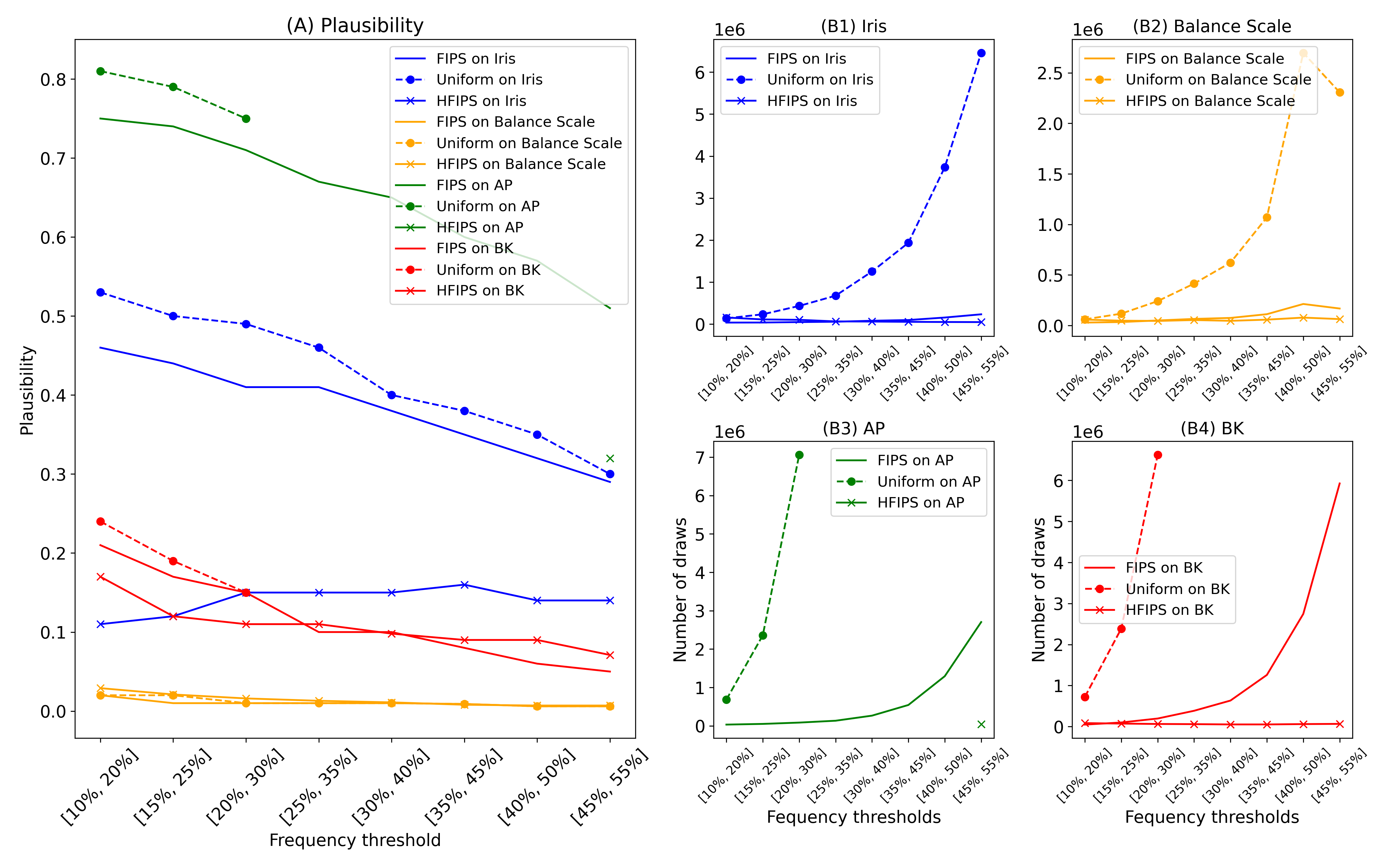}
    \caption{Plausibility evaluation for the \fips{}, \hips{} and \alea{} methods (10 000 sampled patterns) on multiple frequency thresholds}
    \label{fig:plausibility}
\end{figure}

This section addresses Question 5 on the relevance of the patterns  sampled by \fips{}, \hips{} and \alea{} according to the plausibility criterion. We use the protocol introduced by \cite{DBLP:conf/kdd/GionisMMT06} and adapted to numerical data in \cite{DBLP:conf/sdm/GiacomettiS18}.
The idea is to evaluate the relevance of a pattern by comparing its frequency in the original database with its frequency in a randomized database, denoted $\N_{rand}$, in which correlations are broken. {This randomization is performed by repeatedly selecting at random, from the original database, two values of the same attribute from two different objects and swapping them. This process is applied independently for each attribute, and a random number of swaps is performed. A pattern’s relevance increases with the difference between its frequency in the original database and in randomized variants.} Formally, plausibility is defined as : $$Plausibility(K, \N)= \frac{\sum_{\IP_i \in {K}}\sum_{j=1}^{R} (freq(\IP_i,~ \N) - freq(\IP_i,~ \N_{rand}^{j}))} {\sum_{i=1}^{|K|} (R \times freq(\IP_i, \N))}$$ where $R$ is the number of randomised databases. 
%
We defined frequency intervals (see Figure~\ref{fig:plausibility}) and for each interval and method we sampled 10,000 patterns, rejecting those whose frequencies {did not fall within the specified range}.  
The restriction imposed by the frequency intervals aims to simulate the interest of an analyst who rejects both highly frequent patterns (often too general) and very rare patterns (often not representative).  
Considering multiple intervals allow us to simulate several thresholds values considered useful. Sampling was conducted within a time limit of 5 minutes.

\vspace*{0.2cm}

\textbf{Plausibility. } Plausibility tends to decrease as frequency increases, as the more frequent a pattern is, the more likely it is to appear in a randomized database. Since the \fips{} and \hips{} sampling are biased towards frequency and the product of hyper-volume and frequency, respectively, patterns sampled by these methods are expected to have lower plausibility on average than those sampled by \alea{}. Part A of Figure~\ref{fig:plausibility} confirms this observation. As \alea{} produces patterns with the lowest frequencies, their plausibility is the highest. Conversely, \hips{}, which tends to sample the most frequent patterns, leads to patterns with the lowest plausibility. The results also show that as the frequency threshold increases, the plausibility of the patterns sampled by the different methods converges. Note that some methods fail to sample the required number of patterns within the allocated time. For instance, \alea{} fails to sample the required number of patterns with frequencies greater than 30\% for the \textit{AP} and \textit{BK} databases. On the other hand, \hips{} fails to sample the required number of patterns in \textit{AP} with a frequency lower than 45\% within the allocated time.

\vspace*{0.2cm}

\textbf{Number of draws evaluation.} Parts $B_1$, $B_2$, $B_3$ and $B_4$ in Figure~\ref{fig:plausibility} show that across all databases, the number of draws required by \alea{} to reach the target number of patterns is significantly higher than that of \fips{} and \hips{}. For frequency thresholds between 10\% and 35\%, \alea{} requires 0.87 to 10 times more draws than \hips{} on {\it iris}, 1.04 to 7 times more draws on {\it balance-scale}, and 2 to 11 times more draws than \fips{} on both {\it iris} and {\it balance-scale}. This gap increases for higher frequency thresholds (35\% to 45\%), reaching up to 27 times more draws than \fips{} and 35 times more draws than \hips{} on the {\it iris} database. For \fips{}, the difference is even more pronounced on the {\it AP} and {\it BK} databases, where \alea{} requires 15 to 80 times more draws in low frequency thresholds. At higher thresholds, \alea{} fails to sample the required number of patterns in the allocated time. Similarly, for \hips{}, the gap is particularly pronounced on the {\it BK} and {\it iris} databases, where \alea{} requires up to 100 times more draws, and reaches 134 times more on {\it iris}.

\section{Conclusion and Perspectives }

In this paper, we presented two methods for sampling patterns from numerical data. These methods rely on an interval-based representation of the numerical data that preserves all the original information. The first method, \fips{}, samples patterns proportionally to the frequency interestingness measure. 
\hips{} incorporates the hyper-volume measure into the sampling process in order to sample interval patterns proportionally to the product of frequency and hyper-volume. We theoretically proved that both \fips{} and \hips{} sample patterns proportionally to their respective interestingness measures. We experimentally evaluated the quality of the sampled patterns using several criteria, including frequency, hyper-volume, diversity, efficiency, and plausibility. We also evaluate the robustness of our approaches to the long-tail phenomenon.

Increasingly, analysts want to interact with mining systems by specifying, through queries, the types of patterns they want to explore. In this context, a future work is to design a framework for sampling interval patterns that satisfy user preferences and follows the desired distribution. 
This task is particularly challenging since, unlike itemsets, interval patterns are defined across all attributes of the database and include hidden dependencies between intervals. Another direction is to tackle other interestingness measures that are relevant to numerical data, such as density. Finally, exploring the use of sampling in machine learning contexts could represent a promising direction. For example, sampling approaches could help identify meaningful hyper-parameters for supervised learning tasks. By guiding the search toward informative regions of the hyper-parameter space, our approaches may help to reduce computational costs and improve the explainability of the resulting models, which would be a step toward more interpretable machine learning models.

\vspace*{0.2cm}

\noindent
\textbf{Acknowledgement.} This work was supported by the French National Research Agency (ANR) and Region Normandie under grant HAISCoDe, and partly funded by ANR under project ANR-24-CE23-0950.

\appendix
\section{Proof of the $IPH$ function}
\label{appendix:tilProof}

\begin{proof}  
We aim to prove that the $IPH$ function correctly computes the sum of the hyper-volumes of all interval patterns that cover a given object  $g \in \G$. This requires demonstrating that, for any attribute $m \in \M$, the inner term of Equation~\ref{eq:IPH} corresponds to the sum of the lengths of all intervals that contain the value $\val{}{}$ of $g$ on attribute $m$.

Assume that the distinct values of the $m$ attribute are given in a sorted order by:
\[
\mathcal{D}_m = \{d_1, d_2, \dots, d_k\}, \quad \text{with} \quad d_1 < d_2 < \cdots < d_k.
\]

Let $p$ be the index of the highest value in $\mathcal{D}_m$ being less than or equal to  $\val{}{}$:  
\[
p = \max\{i \mid d_i \leq \val{}{}\}.
\]

Then the candidate sets of lower and upper bounds for an attribute $m$ can be defined as follows:  
\[
\I{\val{}{}} = \{d_1, \dots, d_p\}, \qquad \J{\val{}{}} = \{d_p, \dots, d_k\}.
\]

The total interval length $TIL(\val{}{})$ corresponds to the sum of the lengths of all intervals $[a, b]$ such that $a \in I_{g,m}$, $b \in J_{g,m}$, and $a \leq \val{}{} \leq b$. Formally, we write:
\[
\TIL{m}{g} = \sum_{a \in I_{g,m}} \sum_{b \in J_{g,m}} (b - a).
\]

	By substituting with the ordered sets, we have:
		\[
		{\TIL{m}{g}} = \sum_{i=1}^{p} \sum_{j=p}^{k} (d_j - d_i).
		\]

By expanding this double summation, we obtain:
\[
\TIL{m}{g} = \sum_{i=1}^{p} \sum_{j=p}^{k} (d_j - d_i) = \sum_{i=1}^{p} \sum_{j=p}^{k} d_j - \sum_{i=1}^{p} \sum_{j=p}^{k} d_i.
\]

Since the inner sum is independent of $i$ and repeated $p$ times, the first term can be factored as:
\[
\sum_{i=1}^{p} \sum_{j=p}^{k} d_j = p \cdot \sum_{j=p}^{k} d_j,
\]

Similarly, since the inner sum is independent of  $j$ and repeated $k - p + 1$ times, the second term becomes:
\[
\sum_{i=1}^{p} \sum_{j=p}^{k} d_i = (k - p + 1) \cdot \sum_{i=1}^{p} d_i,
\]

	Recognizing that $p = |\I{\val{}{}}|$ and $k-p+1 = |\J{\val{}{}}|$, we can rewrite this as:
		\[
	\TIL{m}{g} = |\I{\val{}{}}| \cdot \sum_{x \in \J{\val{}{}}} x - |\J{\val{}{}}| \cdot \sum_{x \in \I{\val{}{}}} x.
		\]
		
which matches the expression given in Equation~\ref{eq:IPH} for an attribute $m$. Since the hyper-volume of an interval pattern is the product of the lengths of its intervals, applying the product over all attributes leads to the total hyper-volume of all interval patterns covering object $g$.
\end{proof}

\section{Experimental results}
\label{appendix:expe}

\subsection{Pattern Frequencies Evaluation}

\begin{figure}[H]
    \centering
    \begin{minipage}[b]{0.45\linewidth}
        \centering
        \includegraphics[width=\linewidth]{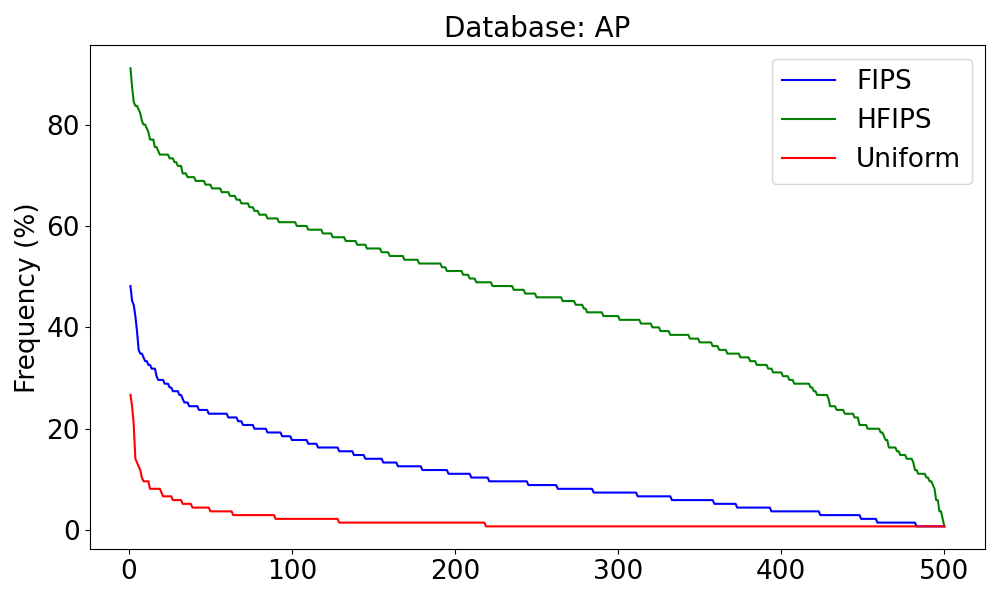}
    \end{minipage}
    \hspace{0.05\linewidth}
    \begin{minipage}[b]{0.45\linewidth}
        \centering
        \includegraphics[width=\linewidth]{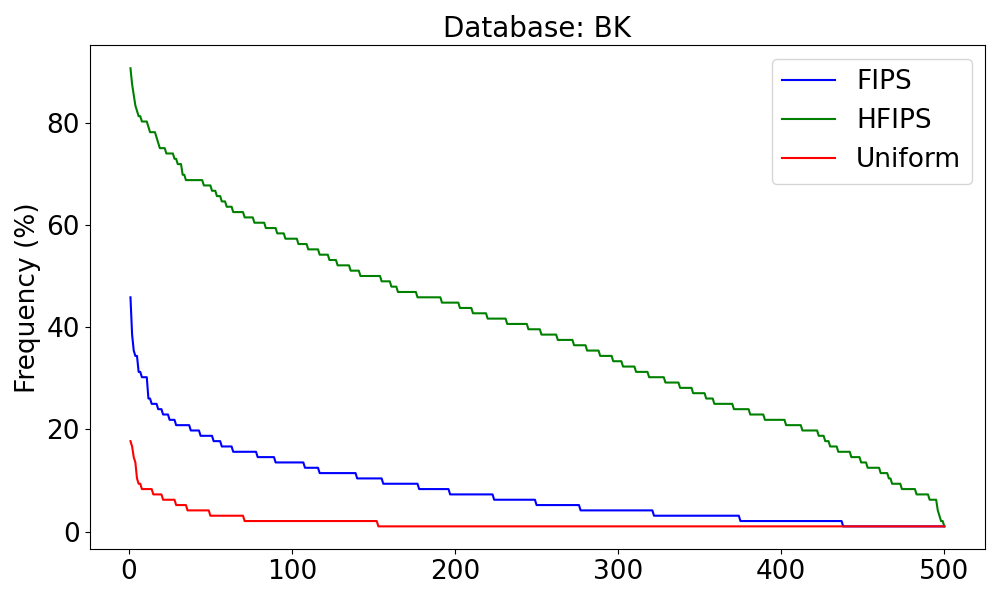}
    \end{minipage}

    \begin{minipage}[b]{0.45\linewidth}
        \centering
        \includegraphics[width=\linewidth]{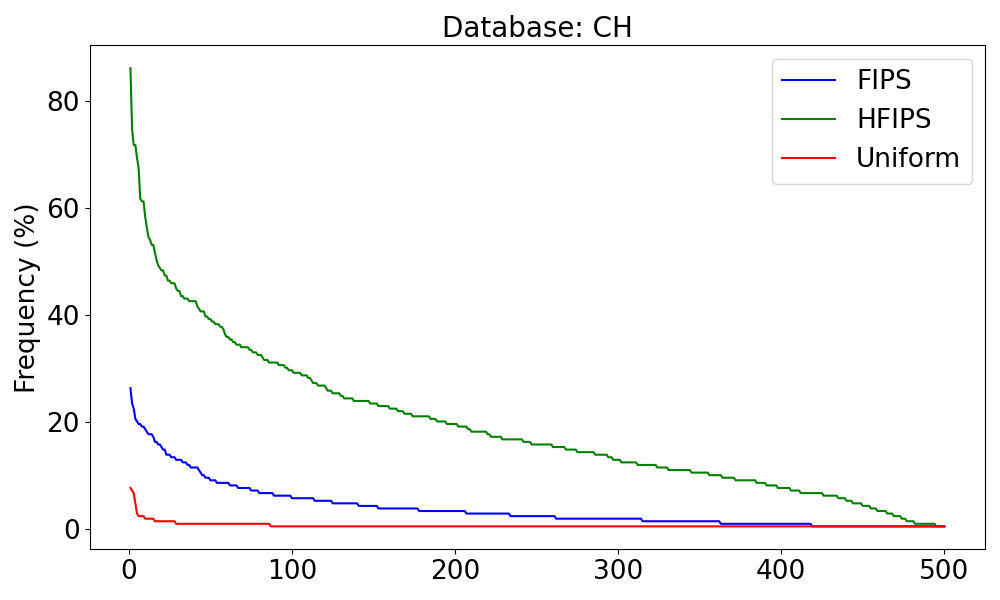}
    \end{minipage}
    \hspace{0.05\linewidth}
    \begin{minipage}[b]{0.45\linewidth}
        \centering
        \includegraphics[width=\linewidth]{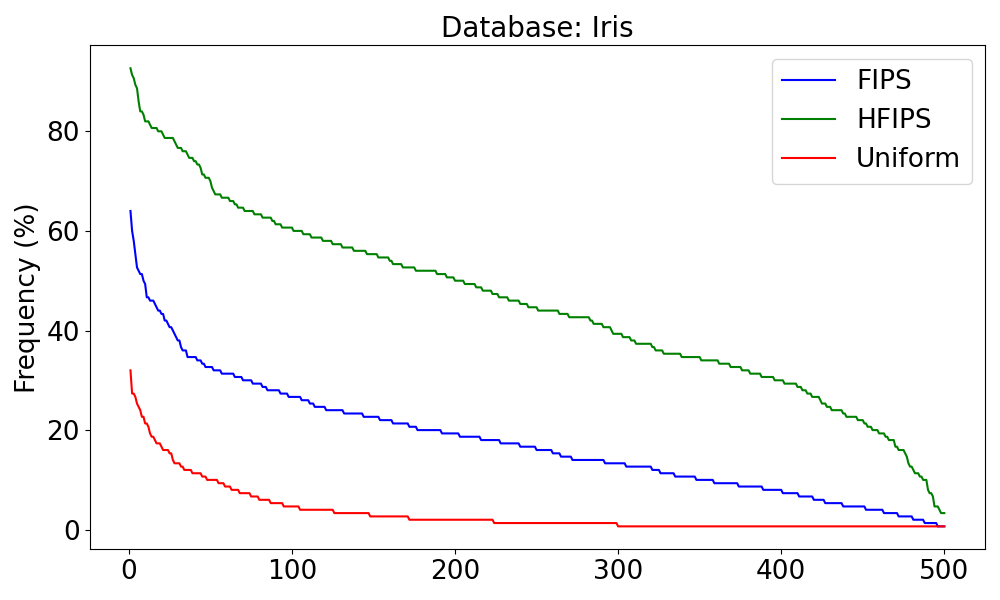}
    \end{minipage}

    \begin{minipage}[b]{0.45\linewidth}
        \centering
        \includegraphics[width=\linewidth]{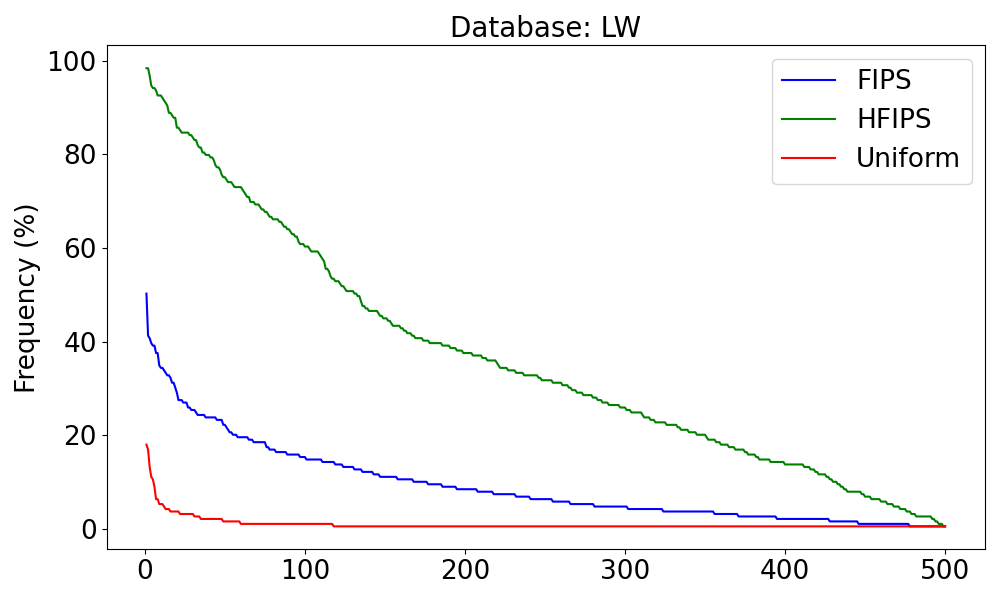}
    \end{minipage}
    \hspace{0.05\linewidth}
    \begin{minipage}[b]{0.45\linewidth}
        \centering
      \includegraphics[width=\linewidth]{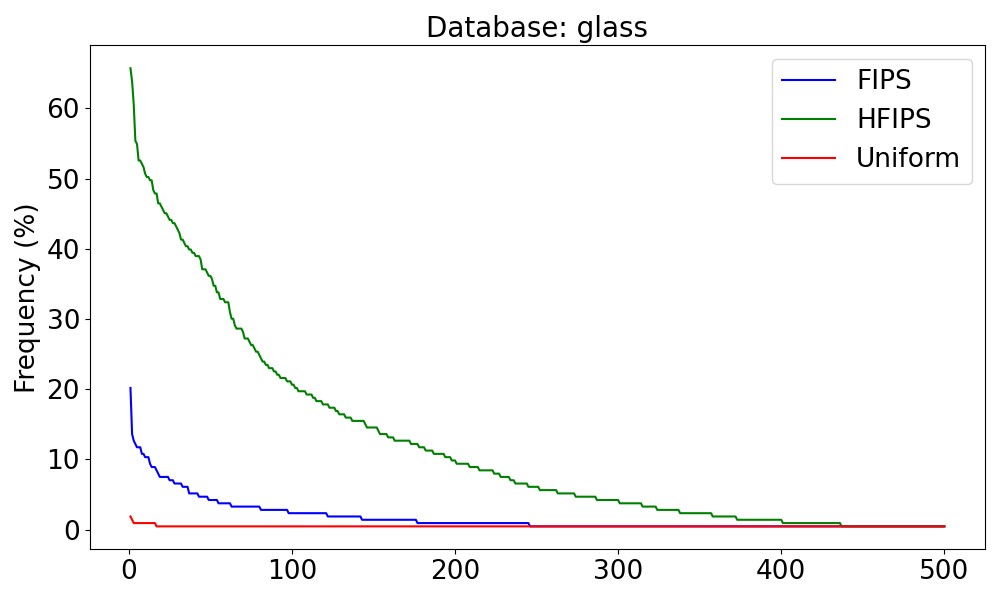}
    \end{minipage}

     \begin{minipage}[b]{0.45\linewidth}
        \centering
        \includegraphics[width=\linewidth]{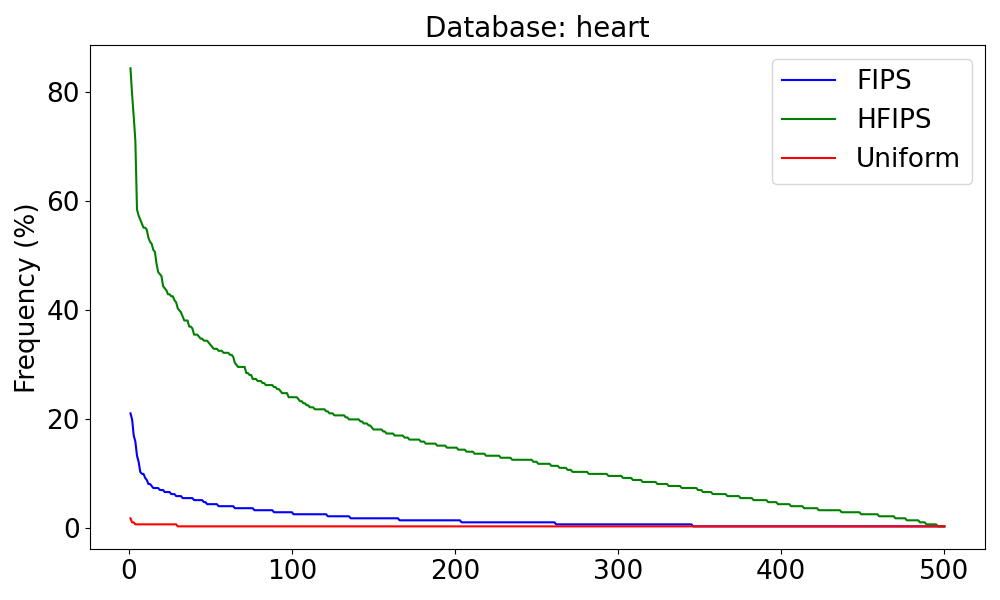}
    \end{minipage}
    \hspace{0.05\linewidth}
    \begin{minipage}[b]{0.45\linewidth}
        \centering
        \includegraphics[width=\linewidth]{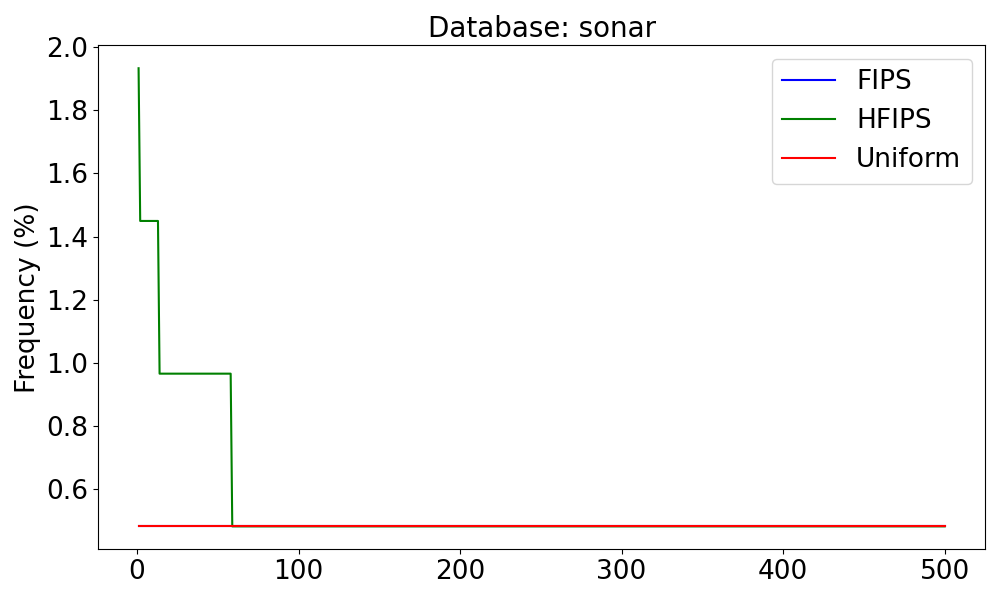}
    \end{minipage}
    \end{figure}
\begin{figure}[H]
\begin{minipage}[b]{0.45\linewidth}
        \centering
        \includegraphics[width=\linewidth]{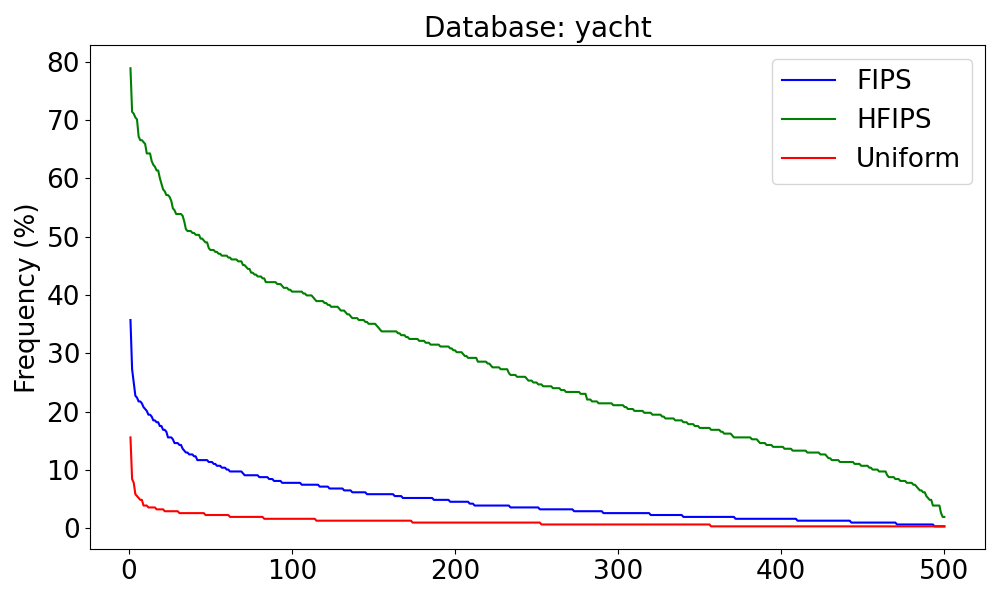}
    \end{minipage}
\end{figure}

\subsection{Pattern Volume Times Frequency Evaluation}

\begin{figure}[H]
    \centering
    \begin{minipage}[b]{0.45\linewidth}
        \centering
        \includegraphics[width=\linewidth]{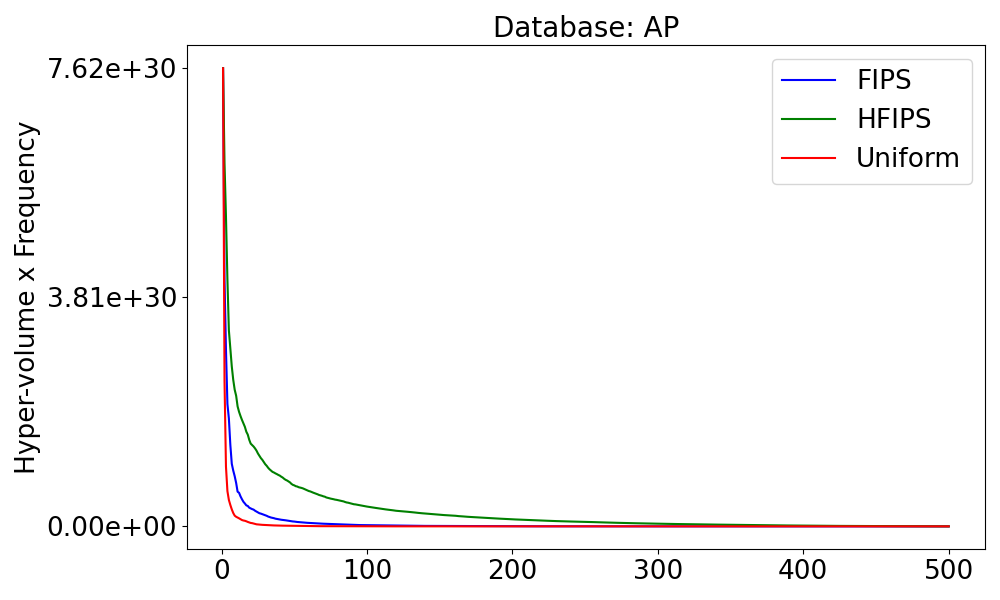}
    \end{minipage}
    \hspace{0.05\linewidth}
    \begin{minipage}[b]{0.45\linewidth}
        \centering
        \includegraphics[width=\linewidth]{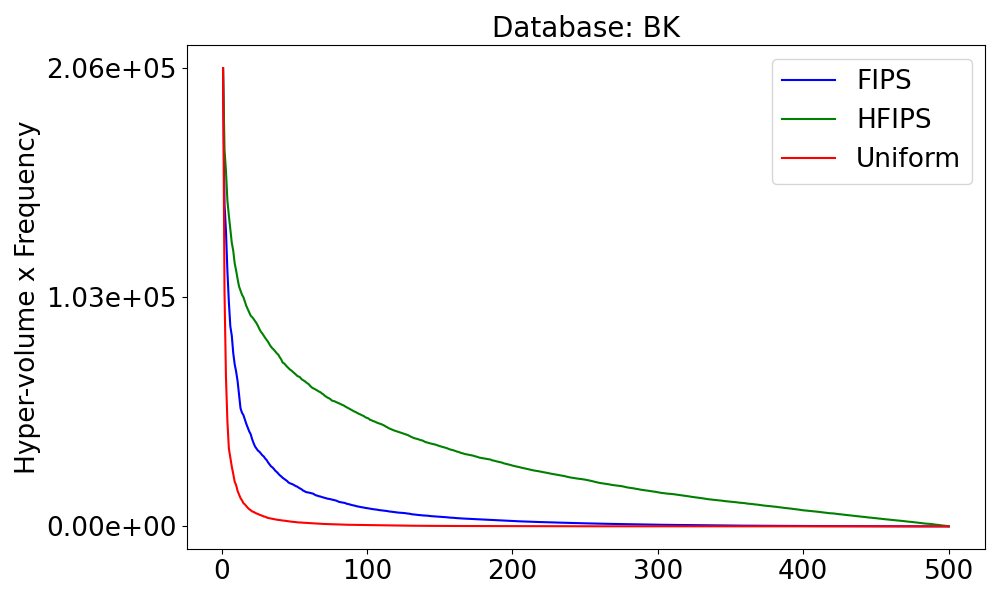}
    \end{minipage}

    \begin{minipage}[b]{0.45\linewidth}
        \centering
        \includegraphics[width=\linewidth]{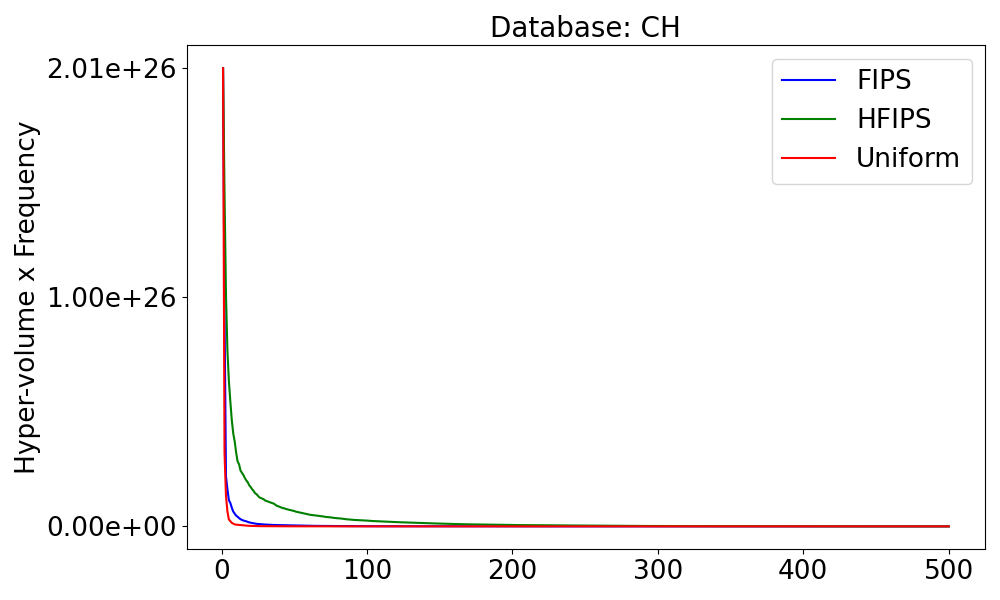}
    \end{minipage}
    \hspace{0.05\linewidth}
    \begin{minipage}[b]{0.45\linewidth}
        \centering
        \includegraphics[width=\linewidth]{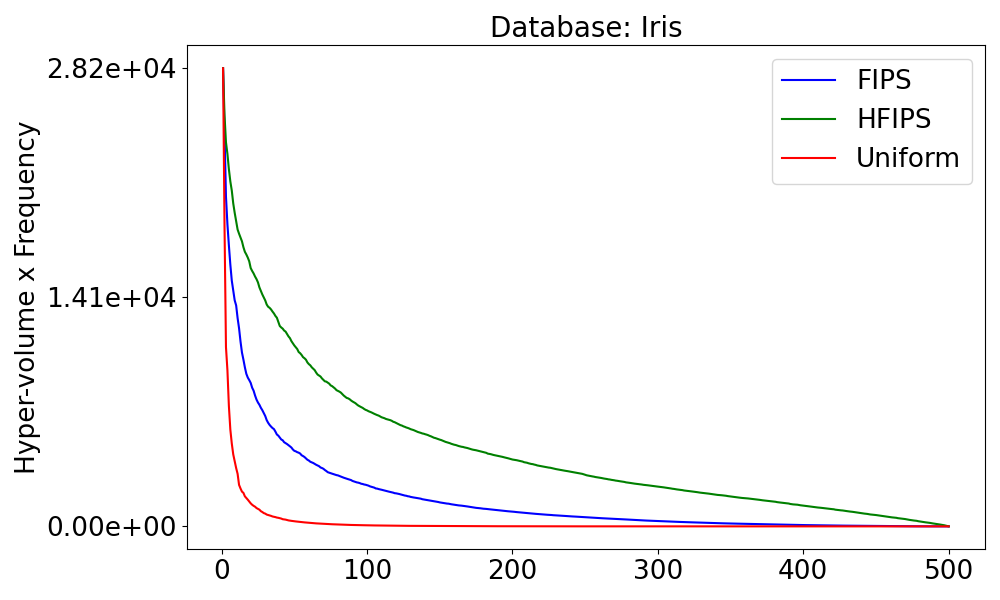}
    \end{minipage}

    \begin{minipage}[b]{0.45\linewidth}
        \centering
        \includegraphics[width=\linewidth]{Images/VolumeTimesFrequency/Volume_evolution_with_hips_2_BK.png}
    \end{minipage}
    \hspace{0.05\linewidth}
    \begin{minipage}[b]{0.45\linewidth}
        \centering
      \includegraphics[width=\linewidth]{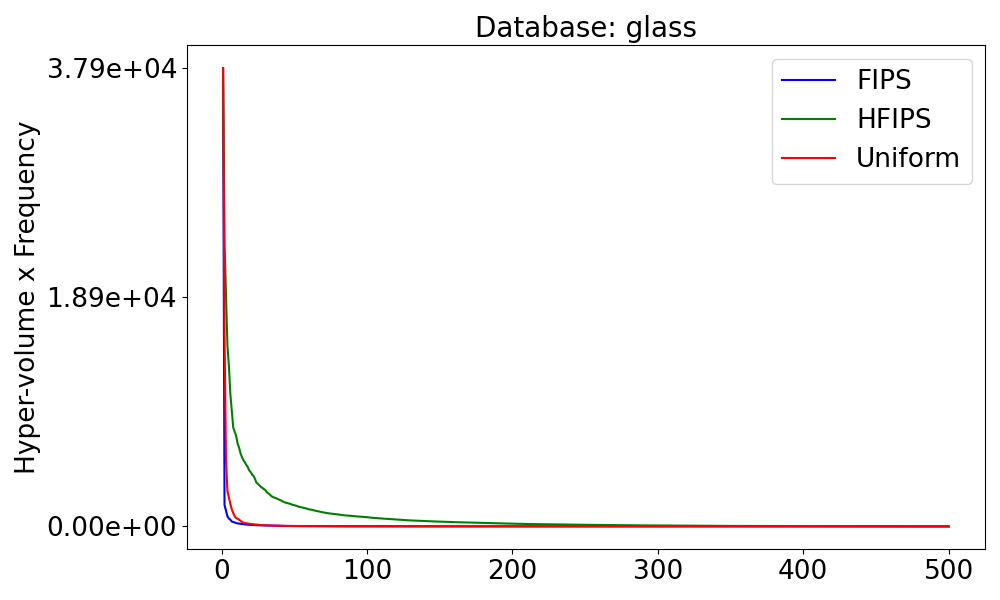}
    \end{minipage}
    \end{figure}
\begin{figure}[H]
     \begin{minipage}[b]{0.45\linewidth}
        \centering
        \includegraphics[width=\linewidth]{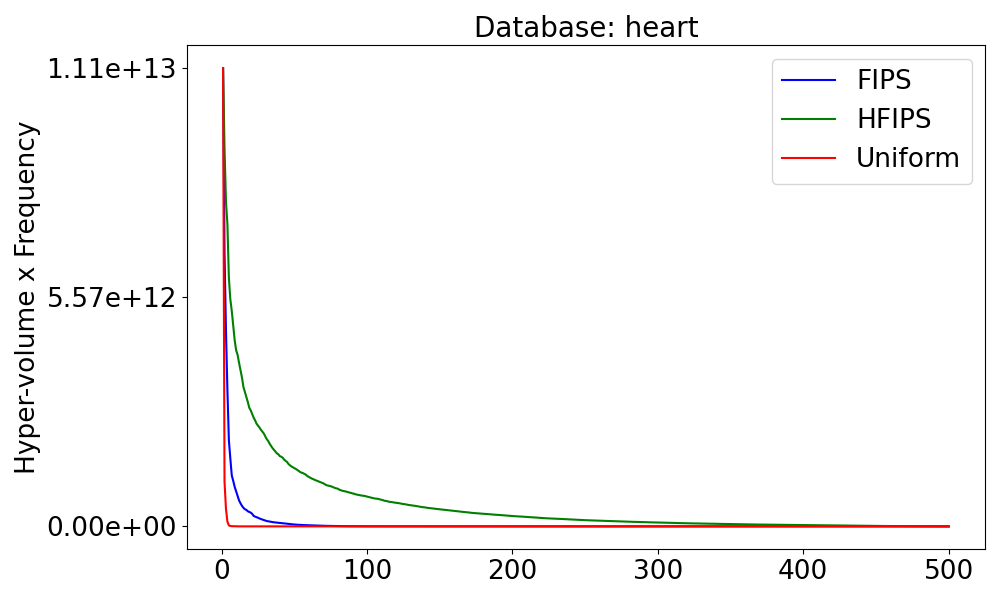}
    \end{minipage}
    \hspace{0.05\linewidth}
    \begin{minipage}[b]{0.45\linewidth}
        \centering
        \includegraphics[width=\linewidth]{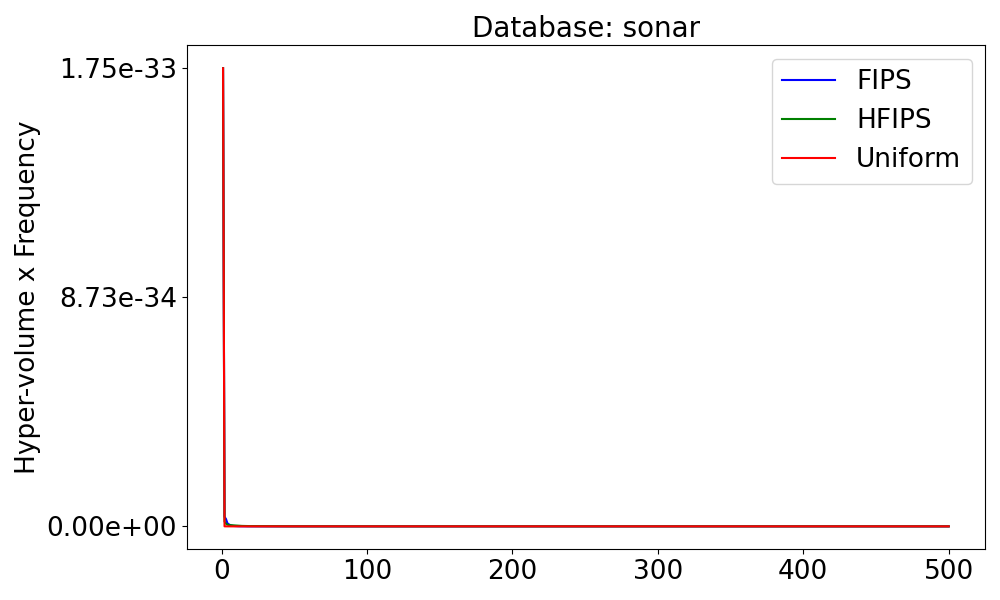}
    \end{minipage}

\begin{minipage}[b]{0.45\linewidth}
        \centering
        \includegraphics[width=\linewidth]{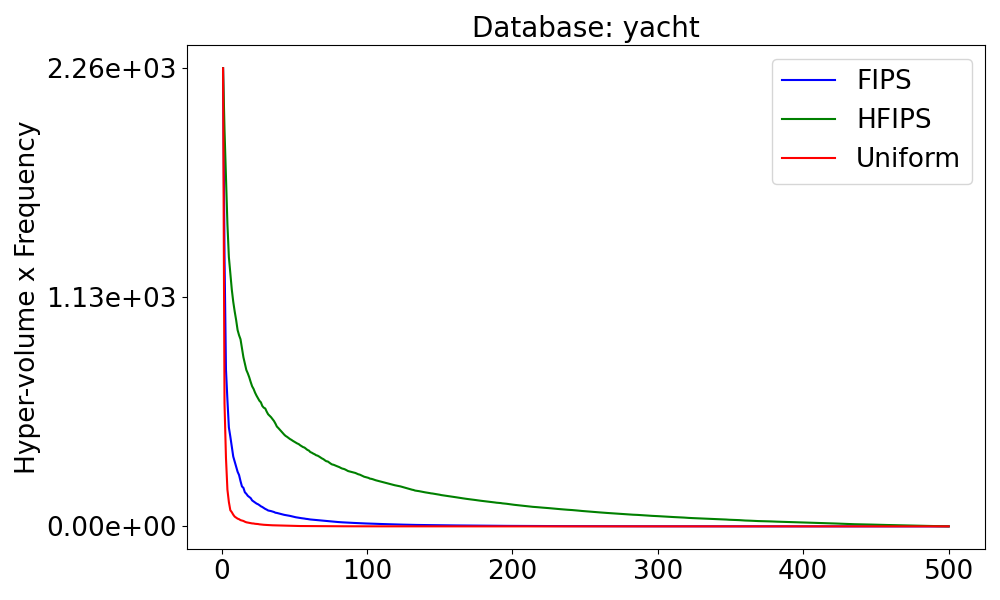}
    \end{minipage}
\end{figure}

\subsection{Jaccard based diversity evaluation}

\begin{figure}[H]
    \centering
    \begin{minipage}[b]{0.45\linewidth}
        \centering
        \includegraphics[width=\linewidth]{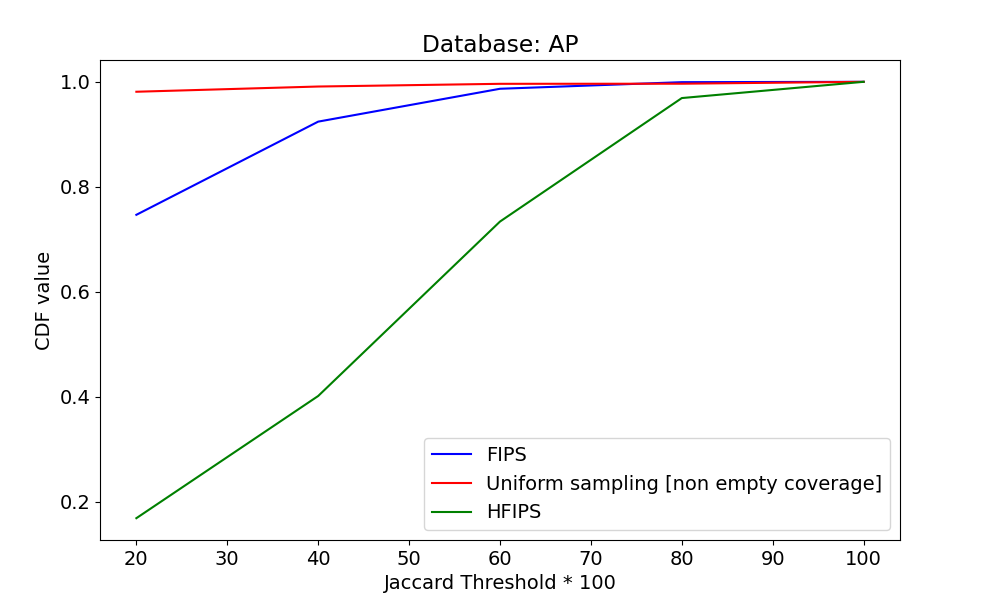}
    \end{minipage}
    \hspace{0.05\linewidth}
    \begin{minipage}[b]{0.45\linewidth}
        \centering
        \includegraphics[width=\linewidth]{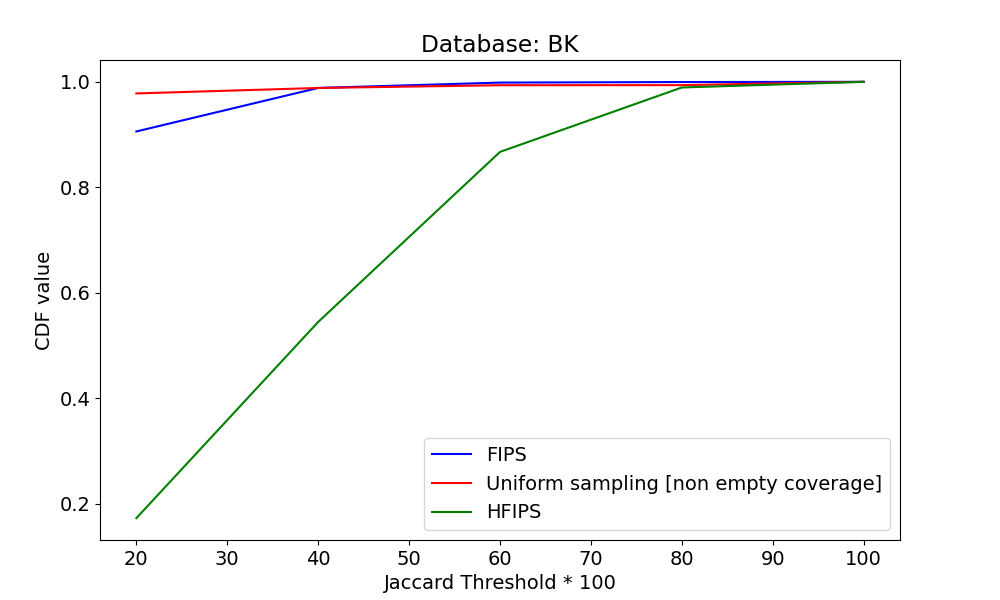}
    \end{minipage}

    \begin{minipage}[b]{0.45\linewidth}
        \centering
        \includegraphics[width=\linewidth]{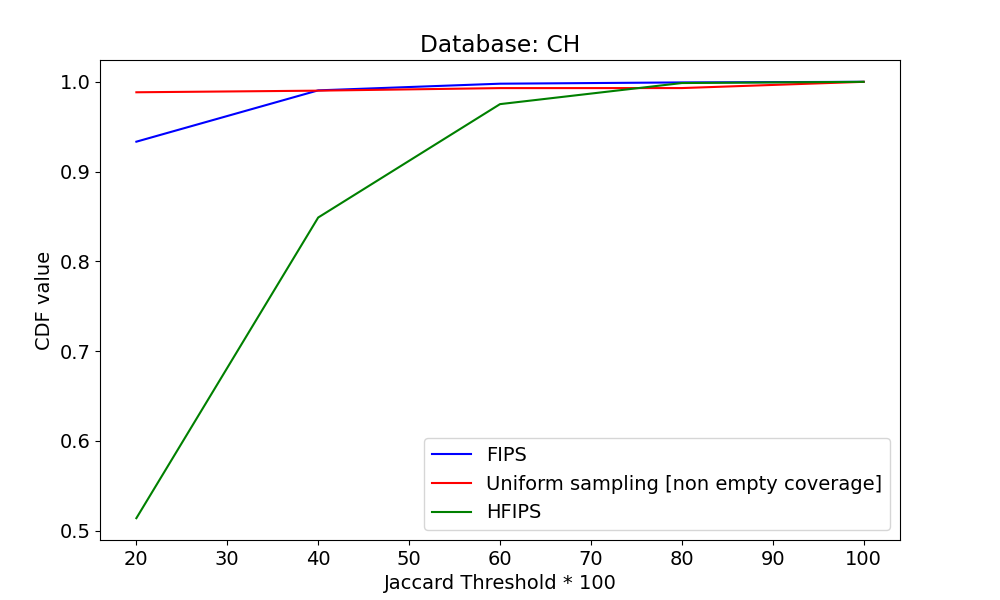}
    \end{minipage}
    \hspace{0.05\linewidth}
    \begin{minipage}[b]{0.45\linewidth}
        \centering
        \includegraphics[width=\linewidth]{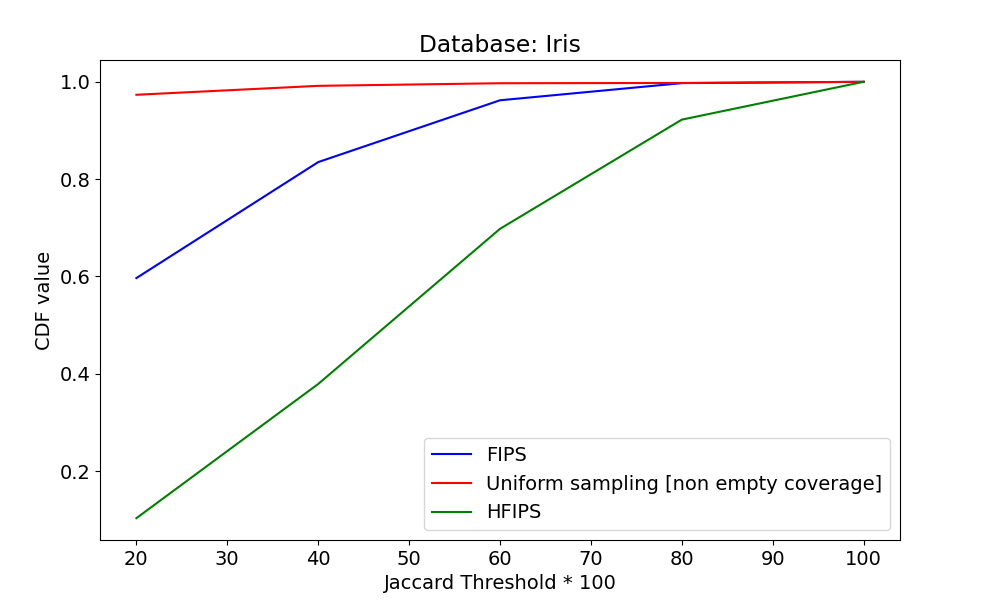}
    \end{minipage}

    \end{figure}
\begin{figure}[H]
    \begin{minipage}[b]{0.45\linewidth}
        \centering
        \includegraphics[width=\linewidth]{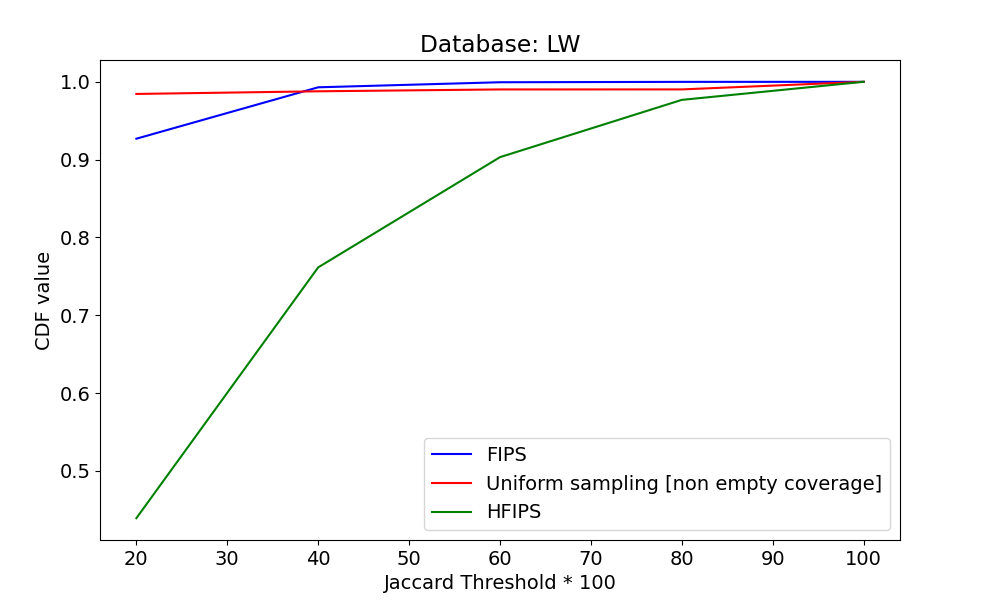}
    \end{minipage}
    \hspace{0.05\linewidth}
    \begin{minipage}[b]{0.45\linewidth}
        \centering
      \includegraphics[width=\linewidth]{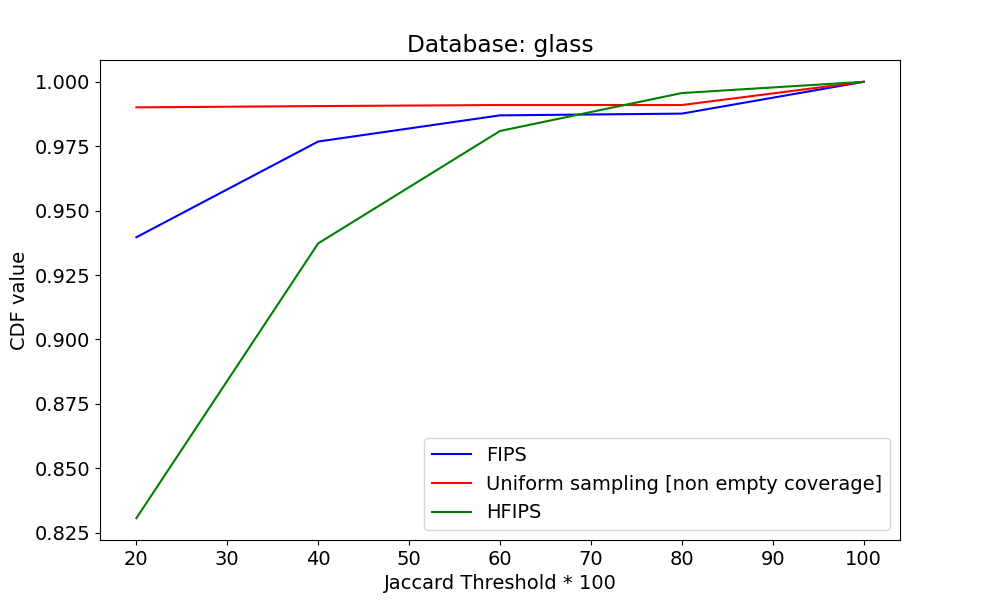}
    \end{minipage}

     \begin{minipage}[b]{0.45\linewidth}
        \centering
        \includegraphics[width=\linewidth]{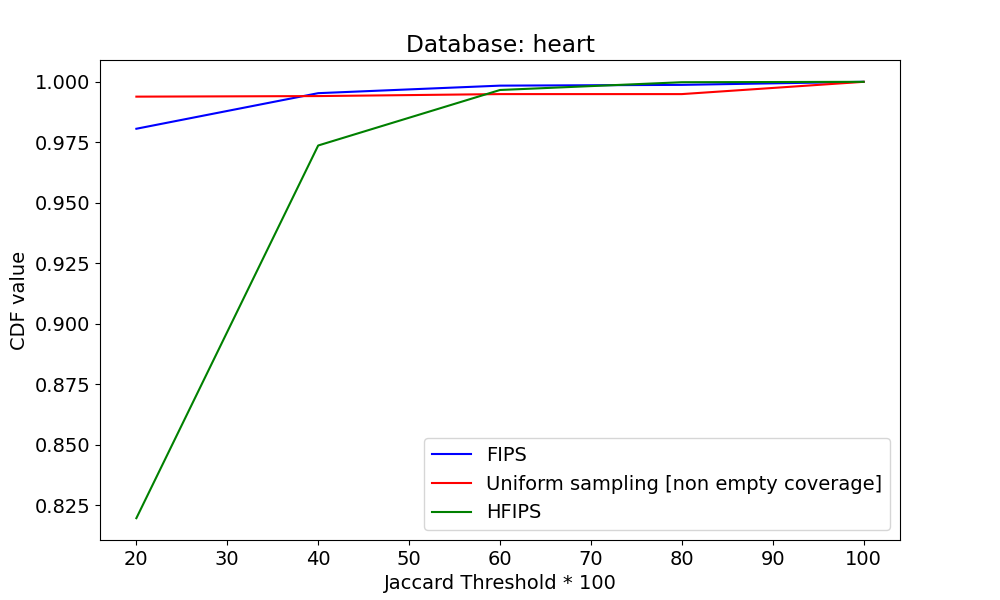}
    \end{minipage}
    \hspace{0.05\linewidth}
    \begin{minipage}[b]{0.45\linewidth}
        \centering
        \includegraphics[width=\linewidth]{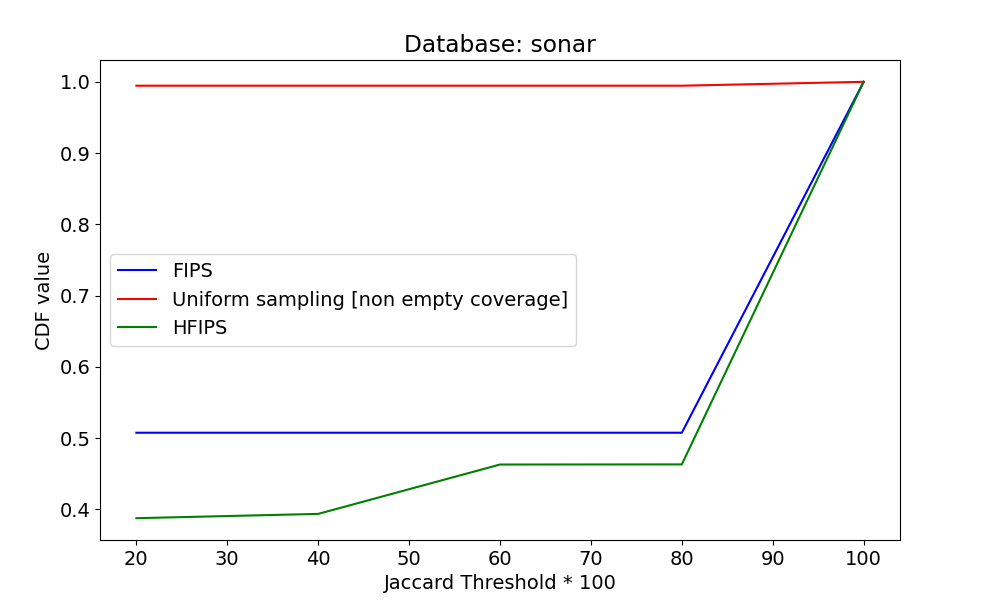}
    \end{minipage}

\begin{minipage}[b]{0.45\linewidth}
        \centering
        \includegraphics[width=\linewidth]{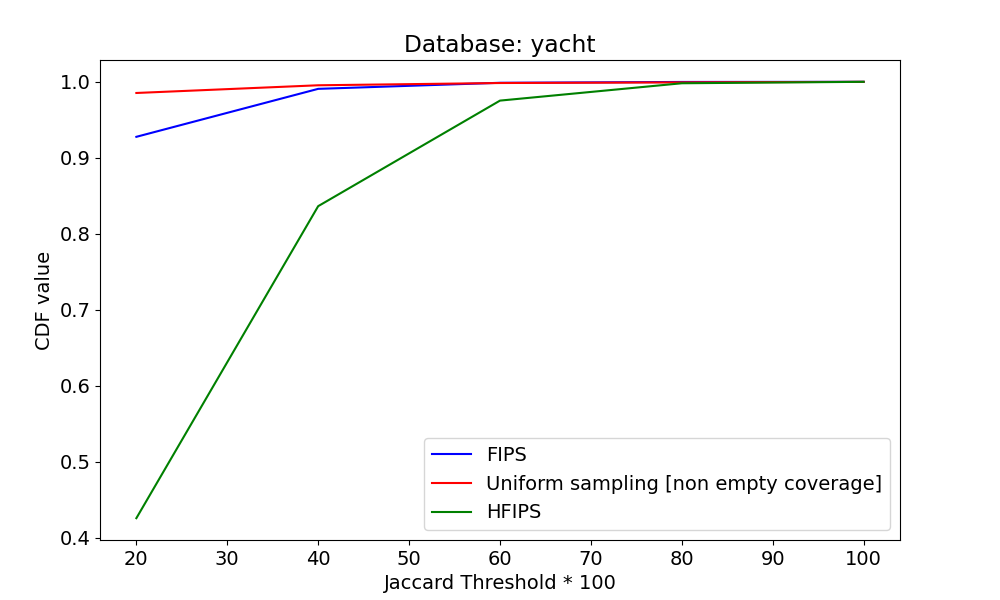}
    \end{minipage}
\end{figure}

\subsection{CPU time evaluation}

\begin{figure}[H]
    \centering
    \begin{minipage}[b]{0.45\linewidth}
        \centering
        \includegraphics[width=\linewidth]{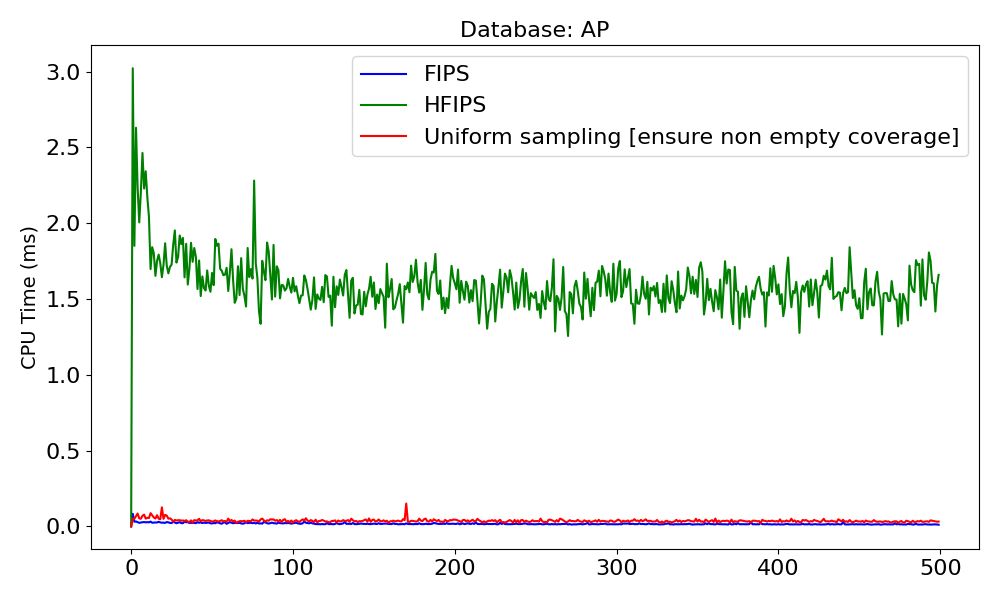}
    \end{minipage}
    \hspace{0.05\linewidth}
    \begin{minipage}[b]{0.45\linewidth}
        \centering
        \includegraphics[width=\linewidth]{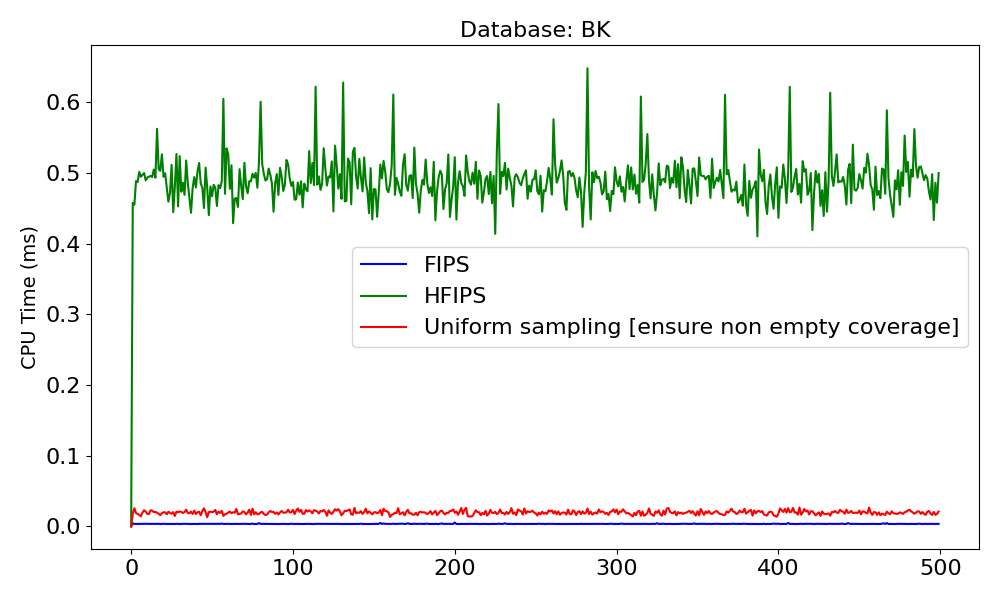}
    \end{minipage}
    \end{figure}
\begin{figure}[H]
    \begin{minipage}[b]{0.45\linewidth}
        \centering
        \includegraphics[width=\linewidth]{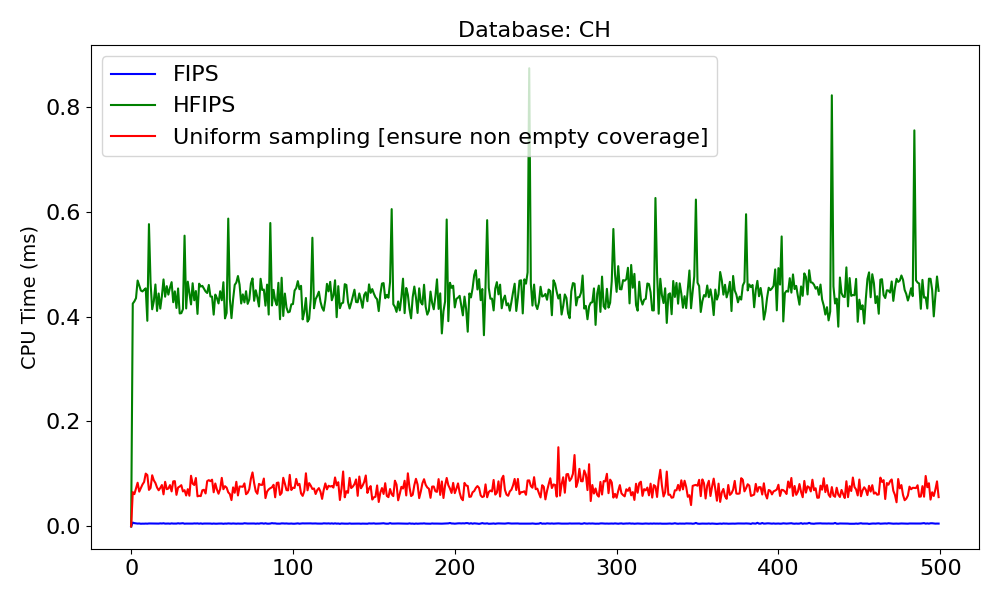}
    \end{minipage}
    \hspace{0.05\linewidth}
    \begin{minipage}[b]{0.45\linewidth}
        \centering
        \includegraphics[width=\linewidth]{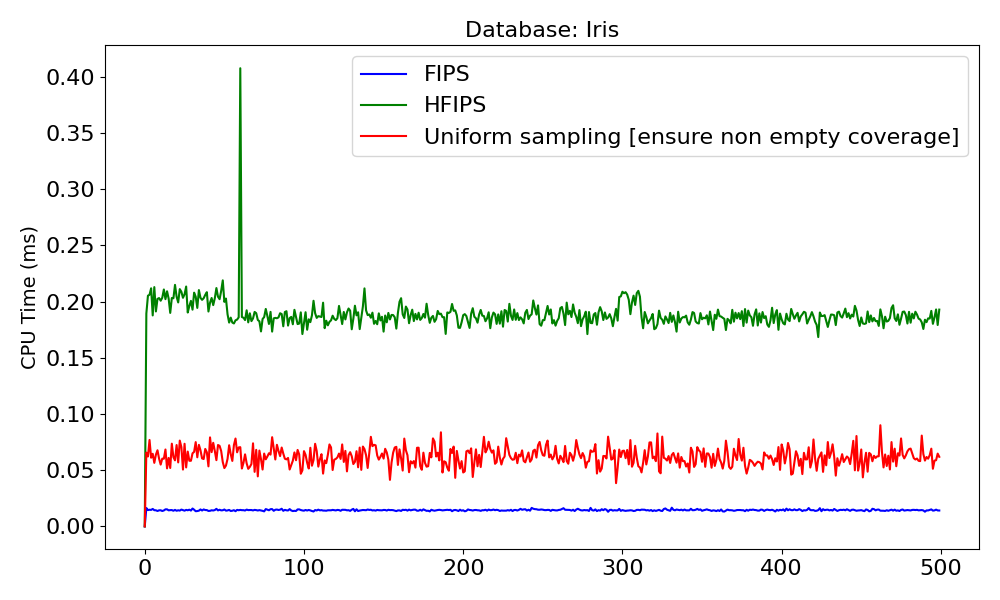}
    \end{minipage}

    \begin{minipage}[b]{0.45\linewidth}
        \centering
        \includegraphics[width=\linewidth]{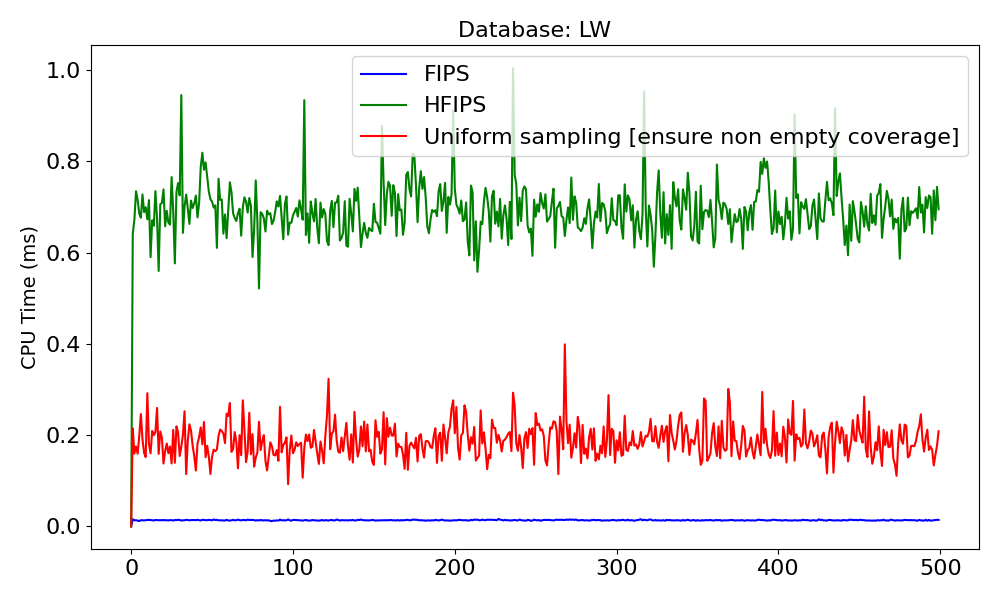}
    \end{minipage}
    \hspace{0.05\linewidth}
    \begin{minipage}[b]{0.45\linewidth}
        \centering
      \includegraphics[width=\linewidth]{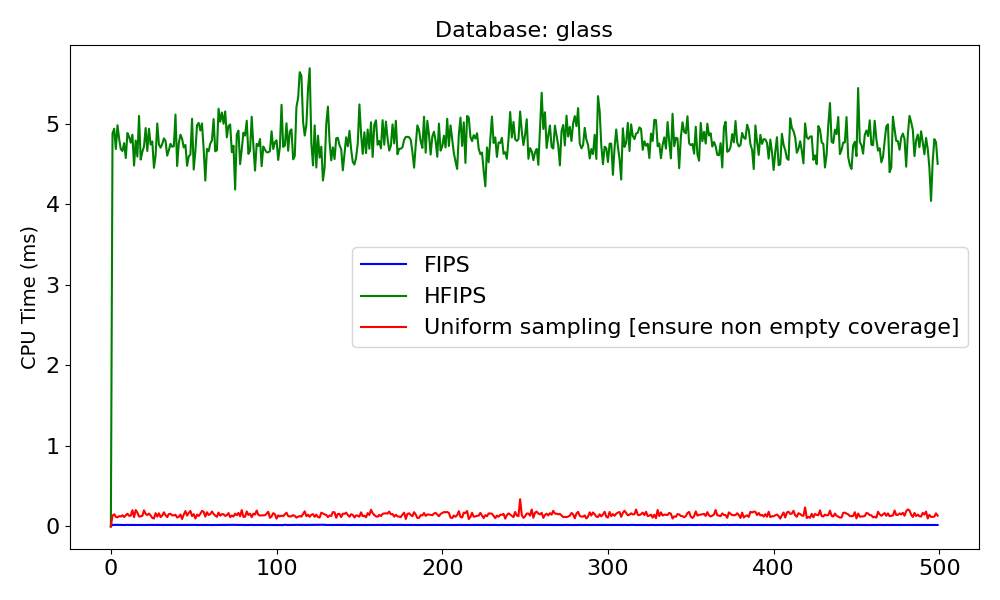}
    \end{minipage}

     \begin{minipage}[b]{0.45\linewidth}
        \centering
        \includegraphics[width=\linewidth]{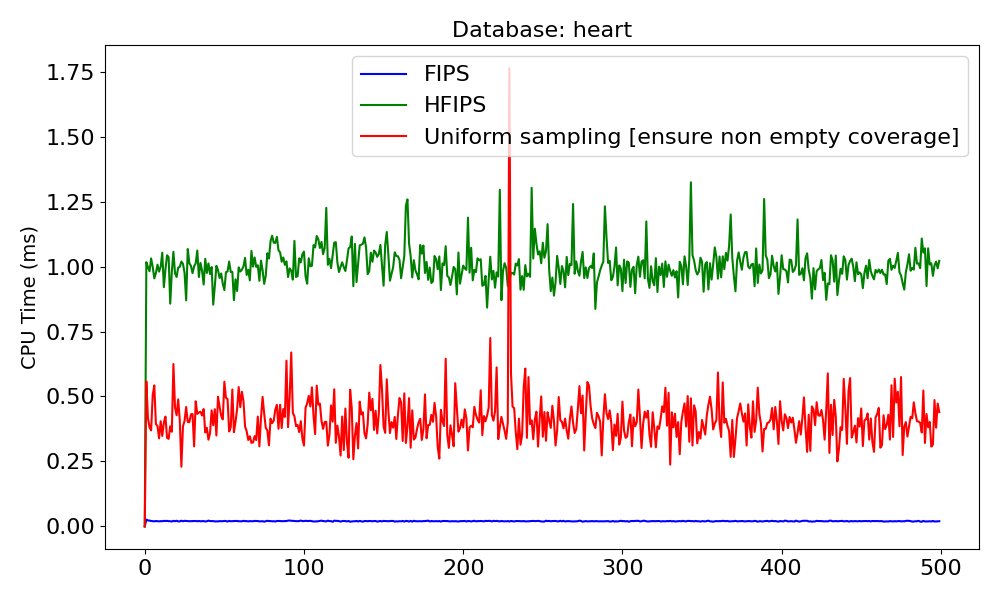}
    \end{minipage}
    \hspace{0.05\linewidth}
    \begin{minipage}[b]{0.45\linewidth}
        \centering
        \includegraphics[width=\linewidth]{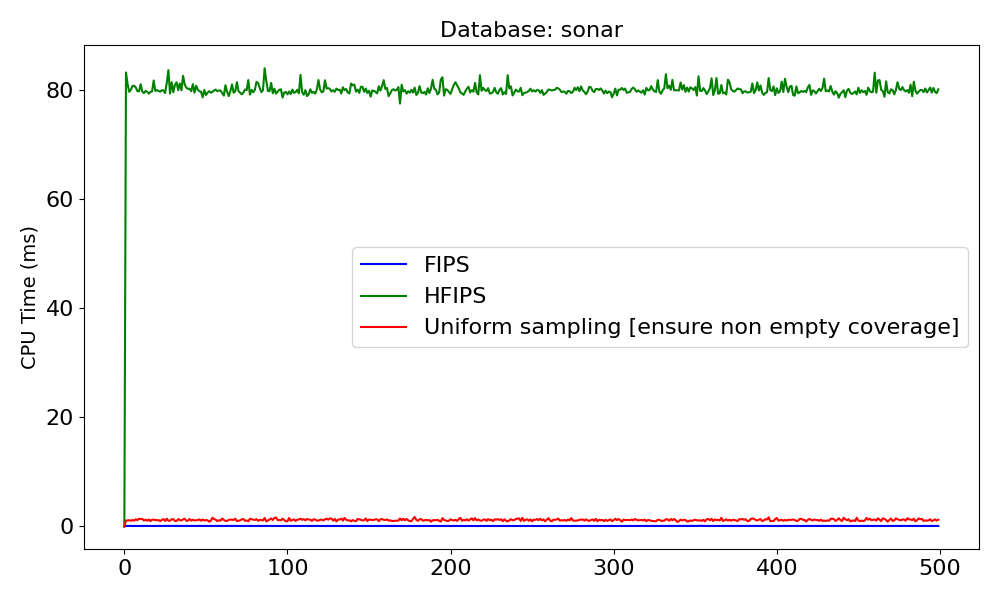}
    \end{minipage}

\begin{minipage}[b]{0.45\linewidth}
        \centering
        \includegraphics[width=\linewidth]{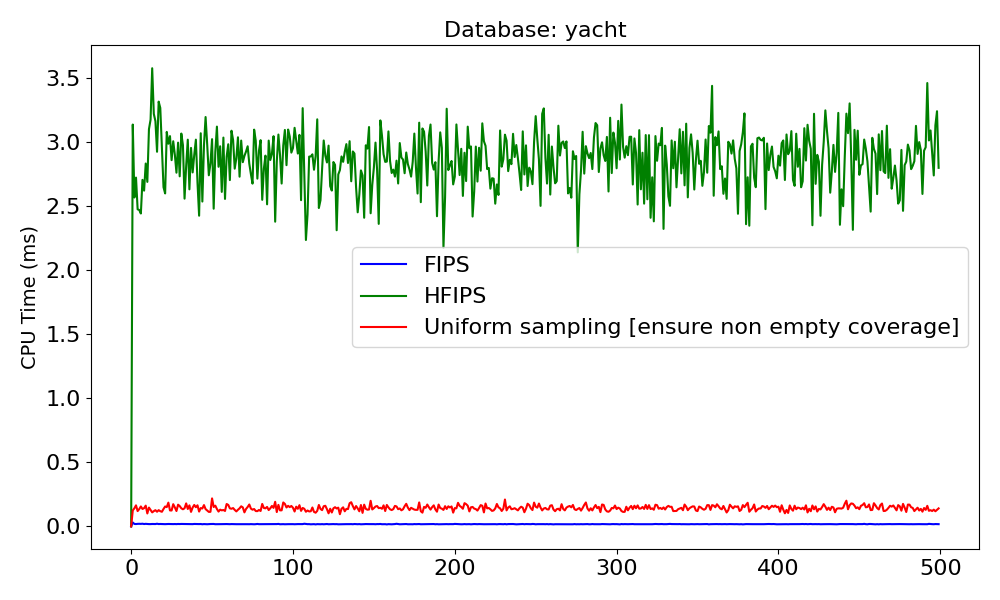}
    \end{minipage}
\end{figure}




\bibliographystyle{elsarticle-num}
\bibliography{biblio}



\end{document}